\documentclass[article,a4paper]{IEEEtran}

\usepackage[noadjust]{cite}
\usepackage[pdftex]{graphicx}

\usepackage{textcomp}
\usepackage{amsmath}
\usepackage{amssymb}
\usepackage{amstext}
\usepackage{braket}
\usepackage{xcolor}
\usepackage{hyperref}
\usepackage{microtype}
\usepackage{gensymb}
\usepackage{comment}
\usepackage{float}
\usepackage{epstopdf}
\usepackage{epsfig}

\begin{document}
\bstctlcite{BSTcontrol}

\title{Structured Light in Turbulence}

\newcommand{\AF}[1]{\textcolor{red}{ #1}}
\newcommand{\MC}[1]{\textcolor{purple}{ #1}}
\newcommand{\NP}[1]{\textcolor{blue}{ #1}}
\newcommand{\IN}[1]{\textcolor{teal}{ #1}}
\newcommand{\Nikki}[1]{\textcolor{violet}{ #1}}

\author{Mitchell~A.~Cox,~\IEEEmembership{Member,~OSA,}
        Nokwazi~Mphuthi,~\IEEEmembership{Member,~OSA,}
        Isaac~Nape,~\IEEEmembership{Member,~OSA,}
        Nikiwe~P.~Mashaba,~\IEEEmembership{Member,~OSA,}
        Ling~Cheng,~\IEEEmembership{Senior~Member,~IEEE,}
        and~Andrew~Forbes,~\IEEEmembership{Fellow,~OSA}
\thanks{N. Mphuthi, I. Nape, N.P. Mashaba and A. Forbes are with The School of Physics, University of the Witwatersrand, Johannesburg, South Africa, e-mail: \href{mailto:andrew.forbes@wits.ac.za}{andrew.forbes@wits.ac.za} .}
\thanks{N. Mphuthi is also with  The South African Radio Astronomy Observatory (SARAO), P. O. Box 443, Krugersdorp 1740, South Africa.}
\thanks{M.A. Cox and L. Cheng are with The School of Electrical and Information Engineering, University of the Witwatersrand, Johannesburg, South Africa.}%
\thanks{N.P. Mashaba is also with the Opto-mechatronics group in the Optronics Sensor Systems department at The Council for Scientific and Industrial Research (CSIR), Defence and Security Cluster, P. O. Box 395, Pretoria, South Africa.}
}


\maketitle

\begin{abstract}
\textbf{Optical communication is an integral part of the modern economy, having all but replaced electronic communication systems.  Future growth in bandwidth appears to be on the horizon using structured light, encoding information into the spatial modes of light, and transmitting them down fibre and free-space, the latter crucial for addressing last mile and digitally disconnected communities. Unfortunately, patterns of light are easily distorted, and in the case of free-space optical communication, turbulence is a significant barrier.  Here we review recent progress in structured light in turbulence, first with a tutorial style summary of the core concepts, before highlighting the present state-of-the-art in the field. We support our review with new experimental studies that reveal which types of structured light are best in turbulence, the behaviour of vector versus scalar light in turbulence, the trade-off of diversity and multiplexing, and how turbulence models can be exploited for enhanced optical signal processing protocols.  This comprehensive treatise will be invaluable to the large communities interested in free-space optical communication with spatial modes of light.}
\end{abstract}


\IEEEpeerreviewmaketitle

\section{Introduction}
\IEEEPARstart{O}{ptical} communication has been an integral part of human society since time immemorial, from early communication by fire beacons for heralding important events, to flag waving in naval and combat situations.  Packing more information into the channel was the obvious next step, with the first multi-level modulation during Napoleonic times and onwards in ever increasing sophistication, laying the foundations for information theory as we know it today.  Over the past 200 years, ``wire'' based solutions have held supreme, from the early days of copper wire communications in 1812, through to optical fibre networks today.  Replacing legacy networks is frightfully expensive, as is laying new fiber networks, and in some instances is not a viable  solution for last mile connectivity nor to bridge the digital divide in areas with an economic or geographical disconnect \cite{lavery2018tackling}.  As such, free-space optical (FSO) communication is enjoying renewed interest \cite{killinger2002free,khalighi2014survey,ghassemlooy2010terrestrial}: it is high-speed, free from licensing and readily deployable, while the no-cloning theorem in the quantum regime, forbidding amplification stages, makes the quadratic loss with distance in free-space very attract when compared to the exponential loss in optical fiber.  Many FSO solutions have already covered 1000s of kilometers in terrestrial links, even with single quantum photons \cite{yin2017satellite}, with exciting high-altitude projects underway by the likes of Facebook, Google and SpaceX.  

In parallel, present fibre optic solutions are rapidly reaching their capacity limit, requiring new degrees of freedom for packing information into light \cite{Richardson1,richardson2013space}.  A popular topic today is to address the impending bandwidth crunch by exploiting the spatial degree of freedom of light, i.e., the spatial modes of light.  Variously referred to as space division multiplexing (SDM) \cite{li2014space} or its special case, mode division multiplexing (MDM) \cite{Berdague1982A}, the central idea is to use the orthogonal set of modes of light as information carriers.  The idea has been around since the 1980s, but gained prominence since the seminal demonstrations with light carrying orbital angular momentum \cite{Gibson2004,Bozinivic2013,Wang1,Willner2015}, and later with other mode sets \cite{zhao2015capacity,Milione2015e, Milione2015, milione2016mode,Trichili2016,trichili2014detection,Xie2016,ahmed2016mode}, as well as with quantum states in the spatial mode basis \cite{Ndagano2018,liu2020multidimensional,de2020experimental,forbes2019quantum,Sit2017,cao2020distribution,cozzolino2019air,cozzolino2019orbital,Mirhosseini2015,mafu2013higher}.  These so-called structured light fields \cite{roadmap} hold tremendous potential, but there is a limitation: spatial modes of light are not impervious to distortion, and in particular they are adversely affected by atmospheric turbulence.  Scattering of one mode into another increases noise, reducing classical channel capacity due to modal cross-talk \cite{JaimeA.Anguita2008,ren2016experimental,krenn2014communication,zhao2016experimental,rodenburg2012a} and reducing security and/or entanglement in a quantum link \cite{Paterson2005,jha2010c,Tyler2009,oamturb,gopaul2007effect,zhang2016experimentally,pors2011transport,goyal2016effect,leonhard2015universal,sorelli2019entanglement}. Mitigating this remains an open challenge that is intensely studied.  

In this article we review the field of structured light in turbulence.  We begin by introducing the core concepts of structured light and turbulence, before moving on to the toolkit needed to simulate and experiment in the field.  Next, we use the toolkit to introduce a collection of results that highlight the present state-of-the-art, in particular addressing what we understand about structured light in turbulence, what is as yet unanswered, in the process resolving some debates in the community.  Finally, we indicate how the turbulence and structured light influences coding and decoding in a FSO communication link, specifically the impact on optical signal processing.  We intend this tutorial style review, which includes new results and perspectives, to be a resource for both new and experienced researchers in the field.

\section{Structured Light}
\label{sec:higherOrderModes}

\noindent From Maxwell's equations, electromagnetic fields in free space, such as laser beams, can be described by the Helmholtz equation, given by
\begin{equation}
\label{eq:HelmholtzEquation}
[\nabla^2 + n^2k^2] \mathbf{E} = 0,
\end{equation}
for electric fields, $\mathbf{E}$. Here, $k = (2\pi)/\lambda$ is the wave number in vacuum and $n$ is the refractive index of the medium in which the field propagates, where $n=1$ in a vacuum. The wavelength is given by $\lambda$ and for visible and near-infrared wavelengths (for example between 400 and 1600~nm) is on the order of hundreds of terahertz. In the case of scalar fields with a uniform polarisation, we can treat the electric field as a scalar field, $U$, a solution to the scalar Helmholtz equation
\begin{equation}
\label{eq:ScalarHelmholtzEquation}
[\nabla^2 + n^2k^2] U(\mathbf{s},z) = 0,
\end{equation}
where $\mathbf{s}$ refers to the transverse coordinates which can be Cartesian where $\mathbf{s}\triangleq(x,y)$ or cylindrical where $\mathbf{s}\triangleq(r,\phi)$ and $z$ refers to the propagation direction of the field. Here we have separated the temporal component from the field and ignored it as we are only interested in its spatial characteristics, and will consider only paraxial solutions where the variation in amplitude in the $z$ direction is slow.

%
%

\subsection{Scalar Modes}
\label{subsec:scalar}

\begin{figure}[tb]
    \centering
	\includegraphics[width=1\linewidth]{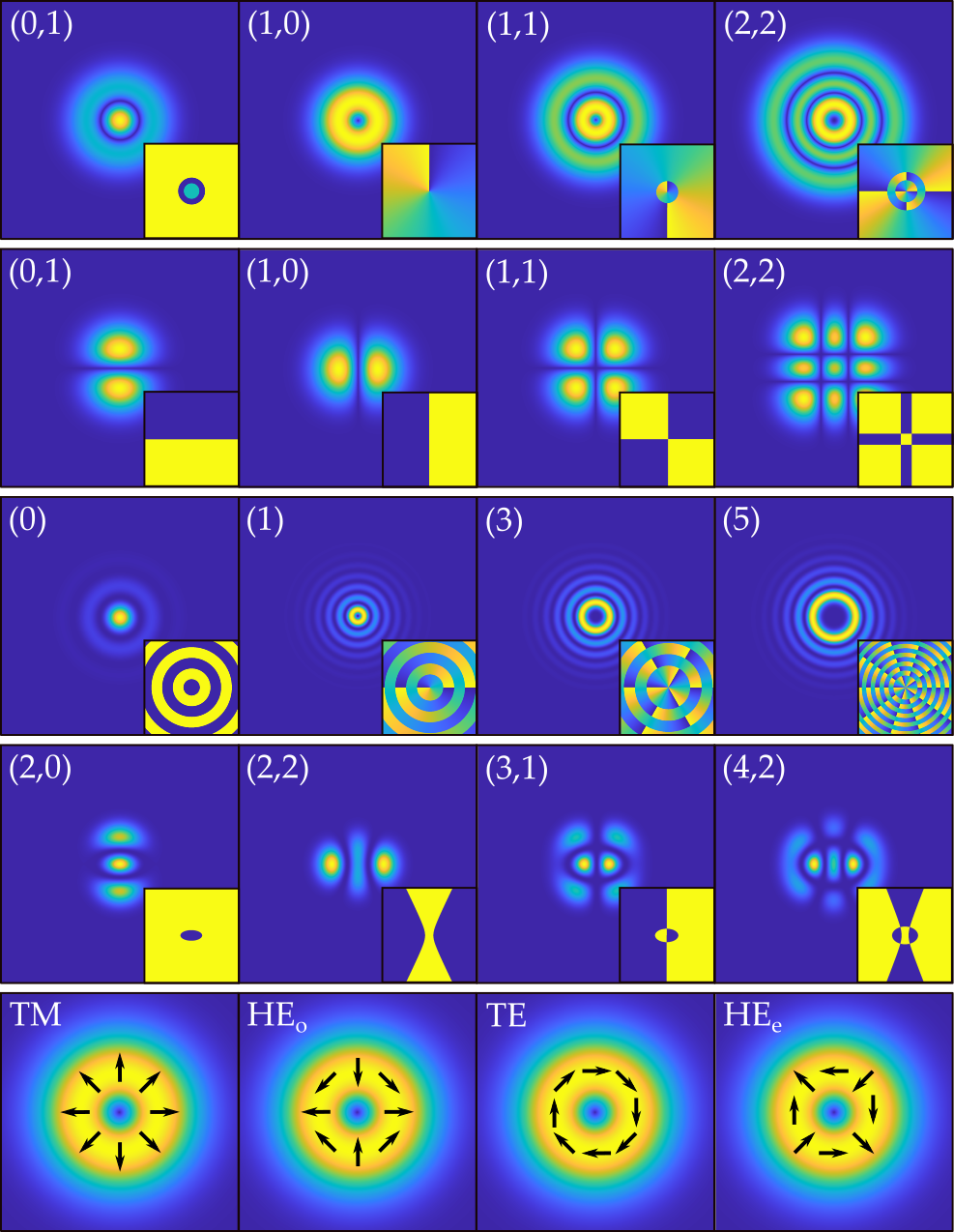}\vspace*{-0.5cm}
	\caption{Intensity patterns with inset phase of scalar and vector modes of various orders. Rows from top to bottom: LG~($\ell,p$), HG~($m,n$), BG~($\ell$), IG$^{\mathrm{e}}$~($p,m,\epsilon=2$) and vector vortex with arrows indicating spatial polarisation direction.}
	\label{fig:scalarmodes} 
\end{figure}

Scalar modes are solutions to the scalar Helmholtz equation and have a spatially homogeneous polarisation. Figure~\ref{fig:scalarmodes} provides illustrative examples of the structured light families described in this section.  In cylindrical coordinates the solutions are known as the Laguerre-Gaussian (LG) modes, given by
\begin{equation}
\label{eq:LG}
\begin{split}
&U(r,\phi,z) = E_0 \left(\sqrt{2}\frac{r}{\omega}\right)^\ell{L_p}^\ell\left(2\frac{r^2}{\omega^2}\right)\frac{w_0}{w(z)}\exp \left[ -i\psi_{p\ell}(z) \right]\times \\
&\exp\left[i\frac{k}{2q(z)}r^2\right]\exp(i\ell\phi),
\end{split}
\end{equation}
where $L_p^\ell(\cdot)$ is the associated Laguerre polynomial, $E_0$ is a constant electric field amplitude, $w(z)$ is the beam size, $w_0$ is the beam size at the beam waist, $z_0=\pi{w_0}^2/\lambda$ is the Rayleigh range, $q(z)=z-i{z_0}$ is the complex beam parameter, $\ell$ is the topological charge of the mode carrying Orbital Angular Momentum (OAM) and $\psi_{p\ell}(z)=(2p+\ell+1)\tan^{-1}(z/z_0)$ is the Gouy phase shift. When $\ell=p=0$, the solution also reduces to the well-known Gaussian beam solution.


In Cartesian coordinates we find the Hermite-Gaussian (HG) modes, given by 
\begin{equation}
\label{eq:HG}
\begin{split}
& U(x,y,z) = E_0 H_m\left(\sqrt{2}\frac{x}{w(z)}\right)H_n\left(\sqrt{2}\frac{y}{w(z)}\right)\frac{w_0}{w(z)}\times \\
& \exp\left[-i\psi_{mn}(z)\right]\exp\left[i\frac{k}{2q(z)}r^2\right],
\end{split}
\end{equation}
where $H_m(\cdot)$ denotes the Hermite polynomials and $\psi_{mn}(z)=(m+n+1)\tan^{-1}(z/z_0)$ is the Gouy phase shift. In this case, the solution reduces to the fundamental Gaussian beam when $m=n=0$. 

There is a third family of exact and orthogonal solutions to the paraxial wave equation known as Ince-Gaussian modes (IG) \cite{bandres2004ince,bandres2004elegant}. These are solutions to the wave equation in elliptical coordinates with spatial patterns described by even and odd Ince polynomials. The Ince-Gaussian modes of order $p$ and degree $m$ are mathematically described as:
\begin{equation}
\label{eq:IG}
\begin{split}
& U(\xi,\eta,z) = D\frac{w_0}{w_z}{C_p}^m(i\xi,\epsilon){C_p}^m(\eta,\epsilon)\exp\left[\frac{-r^2}{w^2(z)}\right]\times \\
& \exp[ikz+\frac{ikr^2}{2R(z)}-i(p+1)\tan^{-1}\left(\frac{z}{zr}\right)],
\end{split}
\end{equation}
where $D$ is a normalisation constant, $\epsilon$ is the ellipticity parameter and ${C_p}^m$ denotes the even Ince polynomial of order $p$ and degree $m$. The variables $\xi$ and $\eta$ are the radial and angular elliptic coordinates respectively, defined as $x=w_0\sqrt{\epsilon/2}\cosh(\xi)\cos(\eta)$ and $y=w_0\sqrt{\epsilon/2}\sinh(\xi)\sin(\eta)$. For odd modes, odd Ince polynomials denoted by ${S_p}^m(\eta,\epsilon)$ are substituted in Eq.~\ref{eq:IG}. The ellipticity parameter of $\epsilon = 0$ and $\epsilon=\infty$ allows a continuous transition between LG and HG modes respectively. For $p=m=0$, the solution reduces to the fundamental Gaussian beam solution.  

Finally, Bessel-Gaussian (BG) modes are another type of solution to the Helmholtz equation in cylindrical coordinates, characterised by a transverse intensity profile based on the family of Bessel functions. These solutions take the general form 
\begin{equation}
\label{eq:BG}
\begin{split}
& U(r,\phi,z) = E_0 \frac{w}{w(z)}\exp[i\psi(z)]\exp\left[i\frac{k}{2q(z)}r^2\right]\times \\
& J_\ell\left(\frac{{\beta}r}{1+i{z/z_0}}\right)\exp\left[\frac{\beta^2z/(2k)}{1+i{z/z_0}}\right]\exp[i\ell\phi],
\end{split}
\end{equation}
where $\beta$ relates to the wave vector of the plane waves which form the mode and $J_\ell$ is the Bessel function of the $\ell^{th}$ order. BG modes carry OAM given by $\ell$. The solution reduces to the fundamental Gaussian beam when $\beta=0$. 

\subsection{Vector Modes}
\label{subsec:vector}
Our discussion on spatial modes has so far been limited to the treatment of the transverse distribution of light as a scalar quantity. However, since spatial modes describe the transverse extent of electric fields, they are vector additions of the  polarisation states, which can be represented by orthogonal unit vectors: the right, $\hat{e}_R = \begin{bmatrix} 1 & 0 \end{bmatrix}^\intercal$, and the left components, $\hat{e}_L = \begin{bmatrix} 0 & 1 \end{bmatrix}^\intercal$, following the Jones formalism. Consequently, we can describe paraxial vector fields in the form
\begin{align}
    \textbf{U}(\textbf(r)) &=\begin{bmatrix}  a \ U_R(\textbf{r}) \\ b \ U_L(\textbf{r})  \end{bmatrix} \\ \nonumber
    &= a \ U_{R}(\textbf{r})\hat{e}_R + b \ U_{L}(\textbf{r})\hat{e}_L,
\end{align}
 where $\textbf{r}$ denotes the coordinates, $a$ and $b$ are complex coefficients while $u_{R,L}(\textbf{r})$ are spatial components of the field and are solutions of the scalar Helmholtz equation.  In general such fields have spatially inhomogeneous polarisation distributions \cite{zhan2009cylindrical,galvez2012poincare,milione2015using}, which can be quantified by a measure of non-separability of the spatial mode and its polarisation \cite{mclaren2015,ndagano2016beam, selyem2019basis}. This class of spatial modes, called vector modes, has received a great deal of attention \cite{rosales2018vectorreview, chen2018vectorial}.




A common form is the special case of cylindrical vector modes, written in the OAM basis as
\begin{equation}
\label{eq:CV}
 \textbf{U}(\textbf{r}) = a \ \psi_{\ell}(\textbf{r})\hat{e}_{L} + b \ \exp(-i \alpha) \psi_{-\ell} (\textbf{r})\hat{e}_{R} .
\end{equation}
Here $\textbf{r} = (r, \phi)$ while $\psi_{\pm \ell} = A(r) \exp(\pm i \ell \phi)$ comprises an enveloping function, $A(r)$, and an azimuthal phase, $\exp(\pm i \ell \phi)$ with topological charge $\pm\ell$ and carrying $\pm \ell \hbar$ of OAM per photon.  The parameters, $a$, $b$ and $\alpha$  control the relative amplitudes and phases between the polarisation components, producing all states on a higher-order Poincare Sphere \cite{milione2011higher}. 
From this we can constructed a four dimensional mode space containing four basis vector modes, i.e.,
\begin{align}
\label{eq:VM}
 \text{TE} \equiv \textbf{U}^{+}_\ell(\textbf{r}) = \frac{i}{\sqrt{2}} \ \big( \psi_{-\ell}(\textbf{r})\hat{e}_{L} - \  \psi_{\ell} (\textbf{r})\hat{e}_{R} \big), \nonumber\\
  \text{TM} \equiv \textbf{U}^{-}_{\ell}(\textbf{r}) = \frac{1}{\sqrt{2}} \big( \ \psi_{-\ell}(\textbf{r})\hat{e}_{L} + \  \psi_{\ell} (\textbf{r})\hat{e}_{R} \big),\nonumber\\
\text{HE}_e \equiv  \textbf{U}^{+}_{-\ell}(\textbf{r}) = \frac{1}{\sqrt{2}} \big(  \ \psi_{\ell}(\textbf{r})\hat{e}_{L} +\  \psi_{-\ell} (\textbf{r})\hat{e}_{R} \big), \nonumber \\
 \text{HE}_o \equiv  \textbf{U}^{+}_{-\ell}(\textbf{r}) = \frac{i}{\sqrt{2}} \big(  \ \psi_{\ell}(\textbf{r})\hat{e}_{L} - \  \psi_{-\ell} (\textbf{r})\hat{e}_{R} \big), \nonumber 
\end{align}
giving the well-known radial, azimuthal and hybrid (even and odd) polarized modes. These modes have become a topic of great interest due to their high dimensional encoding capabilities in quantum and classical communication systems \cite{ndagano2017characterizing,Ndagano2018,nape2018self, willner2018vector, milione2015using,Sit2017}. Examples of the polarisation distribution of their electric fields are illustrated in the last row in Fig.~\ref{fig:scalarmodes}.

Traditional methods of generating vector modes include optical few mode fibers \cite{ramachandran2009generation}, however, more modern methods have adapted controllable laser resonators \cite{naidoo2016controlled}, diffractive optical elements,  \cite{khonina2015generation}, spin-orbit coupling devices utilising liquid-crystals \cite{marrucci2006optical} meta-surfaces \cite{devlin2017spin}, and digital holograms \cite{rosales2017simultaneous}. While they can be detected by way of reciprocity, using the same generation methods, deterministic detection methods have been proposed \cite{ndagano2017deterministic, nagali2010experimental}.

\section{The Atmosphere and Optical Turbulence}
\label{sec:turb}

\subsection{Introductory Theory}
\label{subsec:introtheory}


\noindent When a laser beam propagates through the atmosphere it encounters spatially and temporally varying refractive indices, mainly due to random temperature variations and convective processes. This randomly aberrates the beam wavefront. The Kolmogorov model for turbulent flow is the basis for many contemporary theories of turbulence and is able to relate these temperature fluctuations to refractive index fluctuations. The average size of the turbulent cells are specified by a so-called inner scale, $l_0$, which is typically on the order of millimetres and an outer scale, $L_0$, which is on the order of meters (as a rule of thumb for horizontal free-space laser paths, $L_0$ is approximately the height of the propagating beam above the ground) \cite{Andrews2005}. 

The larger turbulent cells (outer scale) cause a propagating beam to be randomly deflected along its path, resulting in beam wander and angle-of-arrival fluctuations at the receive aperture, with the centroid radial position given by $r_c$. These effects can cause deep fading in a FSO communication system where the beam misses the receiver aperture (in extreme cases) or the detector (focal spot beam wander). The small scale effects distort the wavefront of the beam, resulting in a randomly aberrated phase (and random intensity distribution) at the receiver. Both of these effects are detrimental to structured light.

Models of turbulence typically only provide statistical averages for the random variations of the atmosphere, but in most cases this is sufficient. Kolmogorov turbulence assumes $l_0 = 0$ and $L_0 = \infty$ which simplifies the model significantly and as such, the power spectral density of the refractive index fluctuations given by the Kolmogorov model is described by
\begin{equation}
\Phi^K_n(\kappa) = 0.033 C_{\!n}^2 \kappa ^{-11\!/3} \quad \mathrm{for} \quad  1/L_0 \!\ll\! \kappa \!\ll\! 1/l_0,
\end{equation}
\noindent where $\kappa = 2\pi(f_x \cdot \hat{x} + f_y \cdot \hat{y})$ is the angular spatial frequency vector and $C_{\!n}^2$ is the refractive index structure parameter, which is essentially a measure of the strength of the refractive index fluctuations. 

Values of $C_n^2$ vary from $10^{-17}~\mathrm{m}^{-2/3}$ in ``weak'' turbulence and up to about $10^{-13}~\mathrm{m}^{-2/3}$ in ``strong'' turbulence \cite{Andrews2005}. It is natural to assume that long distance propagation in weak turbulence may be ``just as bad'' as short propagation in strong turbulence. This assumption is indeed true and so the strength of turbulence is in fact denoted by a different parameter, the Rytov variance, which is given by
\begin{equation}
\label{eq:rytovCn2}
\sigma_R^2 = 1.23 C_n^2 k^{7/6} L^{11/6}
\end{equation}
for plane wave propagation in Kolmogorov turbulence. The wave number is given by $k=2\pi/\lambda$ and $L$ is the propagation distance in meters. Typically, weak fluctuations in turbulence are when $\sigma_R^2 \ll 1$, moderate fluctuations when $\sigma_R^2 \approx 1$ and strong fluctuations when $\sigma_R^2 \gg 1$.

There are several more accurate turbulence power spectrum models with increasing complexity, for example the Hill power spectrum (which has no analytical solution), the Tatarskii power spectrum and the von K\'arm\'an power spectrum, given by \cite{Andrews2005}
\begin{equation}
\Phi^{vK}_n(\kappa, l_0, L_0) = 0.033 C_n^2 \frac{\exp{(-\kappa^2/k_m^2)}}{(\kappa^2+k_0^2)^{11/6}} \quad \mathrm{for} \!\quad\! 0 \!\leq\! \kappa \!<\! \infty,
\end{equation}
where $k_m = 5.92/l_0$ and $k_0 = 2\pi/L_0$. A popular power spectrum is the Modified Atmospheric Spectrum, and builds from the von K\'arm\'an and Hill spectrums:
\begin{equation}
\begin{split}
&\Phi^M_n(\kappa, l_0, L_0) = 0.033 C_n^2 \left[ 1 + 1.802\left(\frac{\kappa}{k_l} \right) \right. \\ 
& \left. - 0.254\left(\frac{\kappa}{k_l} \right)^{7/6} \right] \frac{\exp{(-\kappa^2/k_l^2})}{(\kappa^2+k_0^2)^{11/6}}  \quad \mathrm{for} \quad  0 \!\leq\! \kappa \!<\! \infty,
\end{split}
\end{equation}
where $k_l = 3.3/l_0$. We can use these power spectrums to generate individual snapshots of turbulence in the form of phase screens with appropriate statistics, with more detail in Sec.~\ref{subsec:simulatingTurb}.

The Fried parameter, $r_0$, commonly known as the atmospheric coherence length, is a useful alternative to $C_n^2$; for a plane wave (approximately a collimated Gaussian beam) in Kolmogorov turbulence this is given by
\begin{equation}
\label{eq:br0fromCn2}
r_0 = 1.68 \left(C_n^2 L k^2  \right)^{-3/5}
\end{equation}
and more generally for a plane wave in unspecified turbulence,
\begin{equation}
r_0 = \left(0.423 k^2  \int_{0}^{L} C_n^2(z) dz \right)^{-3/5}.
\end{equation}
The atmospheric coherence length is a radius after which the atmospheric turbulence becomes uncorrelated. In other words, in a FSO system, two beams separated by at least $r_0$ will experience uncorrelated, independent fading - an excellent opportunity for channel diversity. 

The Greenwood frequency, $f_G$, is a measure of the rate at which turbulence affects a beam and is easily calculated given a constant wind speed and $r_0$, as $f_G = 0.43(v/r_0)$~[Hz] \cite{Greenwood1977}.

Finally, a common but less descriptive parameter to specify turbulence strength is the Strehl Ratio (SR), which is the ratio of the average on-axis beam intensity with, $\langle I(\mathbf{0}) \rangle$, and without, $I_0(\mathbf{0})$, turbulence and is given by (for a plane wave in Kolmogorov turbulence)
\begin{equation}
\label{eq:bsr}
\mathrm{SR} = \frac{\langle I(\mathbf{0}) \rangle}{I_0(\mathbf{0})} \approx \frac{1}{[1+(D/r_0)^{5/3}]^{6/5}},
\end{equation}
where $D$ is the aperture diameter. In summary, turbulence leads to scintillation, beam wandering and other effects, and is the reason why the on-axis beam intensity, $I(\mathbf{0})$, is reduced on average. The scintillation index, $\sigma_I^2$, is another common parameter similar to the SR, but is useful where knowledge of the intensity in the absence of turbulence is unknown.
\begin{equation}
\label{eq:bscintillationIndex}
\sigma_I^2 = \frac{\langle I^2(\mathbf{0}) \rangle - \langle I(\mathbf{0}) \rangle^2}{\langle I(\mathbf{0}) \rangle^2} = \frac{\langle I^2(\mathbf{0}) \rangle}{\langle I(\mathbf{0}) \rangle^2} - 1
\end{equation}
Similarly to the Rytov index, the scintillation index may be used to roughly characterise turbulence strength. Weak turbulence is when $\sigma_I^2 < 0.3$, medium is when $\sigma_I^2 \approx 1$ and strong is defined as $\sigma_I^2 \gg 1$. For weak fluctuations where $\sigma^2_R \ll 1$, $\sigma^2_R = \sigma^2_I$.

\medskip

In a system making use of multiple spatial modes, the effect of turbulence not only causes attenuation of individual modes, an extension of the SR called Mode Dependent Loss (MDL), but also crosstalk between modes due to aberration of the wavefronts. These phenomena are of significant importance in FSO, with crosstalk being of particular concern due to the ease with which it arises. Polarisation and wavelength, for example, are robust in air because air is neither birefringent nor non-linear, and therefore cannot influence them directly. Spatial phase, on the other hand, is directly affected by atmospheric turbulence which manifests optically as random phase aberrations. 

The precise mechanism and modelling of how structured light interacts with turbulence is very complex and not yet fully understood. Higher order modes have a wavefront that is not well approximated by a plane or spherical wave. Existing models for Gaussian beams are typically inaccurate and so empirical results are usually relied on. There is no ``one size fits all'' solution to the propagation of structured light in turbulence. 

The MDL of an individual mode, $S_i$, is in essence the SR for that particular mode expressed as a loss rather than just a ratio:
\begin{equation}
\label{eq:bmdl}
\mathrm{MDL}_i = 1 - \frac{S_i}{S_{i,0}},
\end{equation}
where $S_{i,0}$ is the intensity of mode $i$ in the absence of turbulence. Typically, the energy in the individual modes is spread to neighbouring modes \cite{Anguita2008a}. This mode crosstalk with respect to mode $i$ is defined as the fraction of the total intensity not in mode $i$:
\begin{equation}
\label{eq:bcrosstalk}
C_i = 1 - \frac{S_i}{\sum_j S_j},
\end{equation}
where $\sum_j S_j$ is the sum of the intensities in all the modes, including mode $i$.


Experimental results have shown that a Johnson-SB fading distribution is a good fit for OAM modes \cite{Anguita2008a,Anguita2009,Funes2015}, however, there has been no work identifying how the (numerous) parameters of the Johnson-SB distribution map to standard variables such as $C_n^2$ or $r_0$. There has been no experimental work determining the fading distributions in real atmosphere for other higher-order mode sets such as the HG, IG or even LG modes where $p\neq0$. 

There has also been some work to characterise and model the crosstalk behaviours of OAM modes \cite{Tyler2009,ren2013,Zhou2015,Zhou2016}.  Analytically, the ensemble average, $\braket{P_j}$ of the normalised power in an OAM mode specified by $\ell_j$ when mode $\ell_i$ is transmitted is given by \cite{Tyler2009}
\begin{equation}
\label{eq:oamCrosstalk}
\braket{P_j} = 
\begin{cases}
1-1.01\!\!\left(\frac{D}{r_0} \right)^{5/3} 	&\text{for } \Delta=0\\
0.142 \frac{\Gamma(\Delta - 5/6)}{\Gamma(\Delta + 11/6)}\!\!\left(\frac{D}{r_0} \right)^{5/3}    	& \text{otherwise}
\end{cases}
\end{equation}
where $\Delta = |i-j|$ and $\Gamma(\cdot)$ is the gamma function. Equation~\ref{eq:oamCrosstalk} has been verified \cite{Rodenburg2012}, but recently was shown to be inaccurate over long distances (1.7~km), probably due to beam wander \cite{Lavery2017}. 

While OAM has been well studied, there are several different modal bases that could be used for FSO and it is coming to light that each set of spatial modes may require a tailored model \cite{krenn2019turbulence}. This, as well as the resilience of different spatial modes is discussed further in Sec.~\ref{sec:robustness}.

\medskip

Beam wander and angle-of-arrival fluctuations are well modelled processes in atmospheric turbulence. These effects have a direct influence on modal crosstalk. A good understanding of turbulence-induced beam wander may be useful in understanding (and perhaps predicting - see Sec.~\ref{sec:memory}) MDL and crosstalk to a certain degree.

\begin{figure}[tb]
    \centering
	\includegraphics[width=1\linewidth]{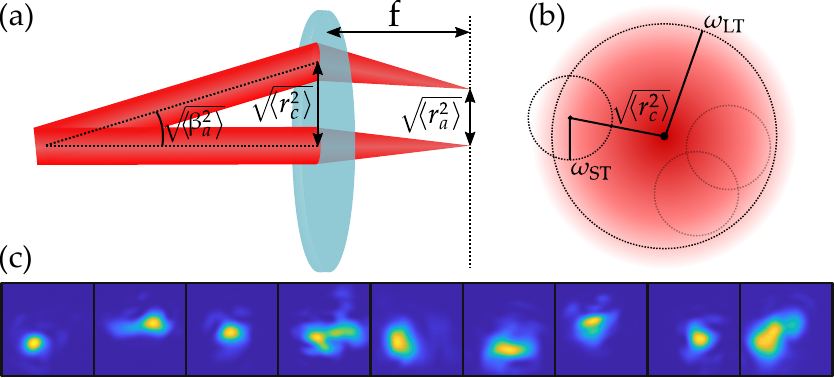}\vspace*{-0.5cm}
	\caption{\label{fig:wanderParams} (a) Diagram of a receive lens with focal length $f$, showing RMS beam wander, angle-of-arrival and focal spot wander parameters. (b) Illustration of the relationship between short- and long-term beam wander parameters. (c) Images of short term beams displaying focal spot beam wander and scintillation.}
\end{figure}
Beam wander at the receiver plane is statistically modelled by a Gaussian distribution which is characterised by a long-term average radial variance, $\braket{r_c^2}$, as well as the size of the received beam without any averaging (the short-term beam), denoted by $\omega_{\mathrm{ST}}$, illustrated in Fig.~\ref{fig:wanderParams}~(b) \cite{Andrews2005}. It is intuitive to think that a smaller short-term beam which wanders will result in an average larger Gaussian shaped long-term beam, according to the central limit theorem \cite{Fante1975}. The size of the long-term beam, $\omega_\mathrm{LT}$, is given by 
\begin{equation}
    \omega_{\mathrm{LT}}^2 = \omega_{\mathrm{ST}}^2 + \braket{r_c^2}.
\end{equation} 
Several analytical solutions for the beam wander radial variance exist. 
For infinite outer scale Kolmogorov turbulence, the beam wander radial variance for a collimated beam is given by \cite{churnside1987aoa}
\begin{equation}
    \label{eq:beamWanderVarInfOuterScaleColl}
    \braket{r_c^2} = 2.42 C_n^2 L^3 \omega_0^{-1/3}.
\end{equation}
Similarly, the variance of angle-of-arrival fluctuations at the receive aperture are given by
\begin{equation}
    \label{eq:angleOfArrivalVariance}
    \braket{\beta_a^2} = 2.91 C_n^2 L (2\omega_G)^{-1/3},
\end{equation}
where $\omega_G$ is the radius of the receive aperture. 
%
%

The detected OAM spectrum due to tilt, $\beta_a$, and lateral displacement, $r_c$, (i.e., misalignment) impinging on a detection hologram (for instance) has been modelled \cite{Vasnetsov2005,Lin2010oamtilt}. This work can be applied to the case of turbulence-induced angle-of-arrival fluctuations and beam wander since the geometry of the receiver system is known. In a system with both tilt and lateral displacement, there is a third parameter which describes what could be called ``swing''. The expression for this combination is long and can be found in Ref.~\cite{Vasnetsov2005}, Eq. 16.

\subsection{Simulating turbulence in the laboratory}
\label{subsec:simulatingTurb}


Here we outline two general approaches to simulating optical turbulence using Spatial Light Modulators (SLMs) and Digital Micromirror Devices (DMDs) for use in a lab environment. They have been utilised in a myriad of experiments for studying the evolution of spatial modes through turbulence \cite{malik2012influence, Zhao2012, cheng2016channel}. The turbulence screens can also be used for computer simulations. 

Various numerical methods have been developed to approximate the phase that is imparted on optical beams upon traversing turbulence. One such technique approximates the phase screen as a random phase function with a variance of \cite{mcglamery1967restoration}
\begin{equation}
\sigma^2(x,y) = \left(\frac{2\pi}{ N\Delta x}\right)^2 \Phi(k_x, k_y),
\end{equation}
for an $N \times N$ grid over which the screen is generated. Here $\Delta x$ is the grid spacing, while $k_x$ and $k_y$ are the spatial frequencies over the grid and $\Phi(k_x, k_y)$ is the phase spectrum given by
\begin{equation}
\Phi(k_x, k_y) = 2\pi k_0^2 L \Phi_n(k_x, k_y).
\end{equation}
Here $\Phi_n(k_x, k_y)$ is the refractive index power spectrum, $L$ is the distance over which the screen is simulated for and $k_0$ is the wave-number (in vacuum) of the optical beam to be used. The desired phase screen can be computed from 
\begin{equation}
  \phi(x,y) =  \text{Real}\left\{ \mathcal{F}^{-1}( \ \chi(k_x, k_y)\sigma(k_x, k_y) \ ) \right\},
  \label{fig:fourierph}
\end{equation}
where $\chi(k_x, k_y)$ is a complex matrix of size $N\times N$ with random entries sampled from a normal distribution with 0 mean and a variance of 1. $\mathcal{F}^{-1}$ denotes the inverse Fourier transform, where the Fast Fourier Transform (FFT) may be used. Although here we show that the screen is extracted from the real part, it can in principal also be extracted from the imaginary part.
Furthermore, the phase spectrum must be set to zero at the origin $\Phi(0, 0)=0$ to avoid a large piston component in the final screen. 

\begin{figure}[tb]
    \centering
    \includegraphics[width=1\linewidth]{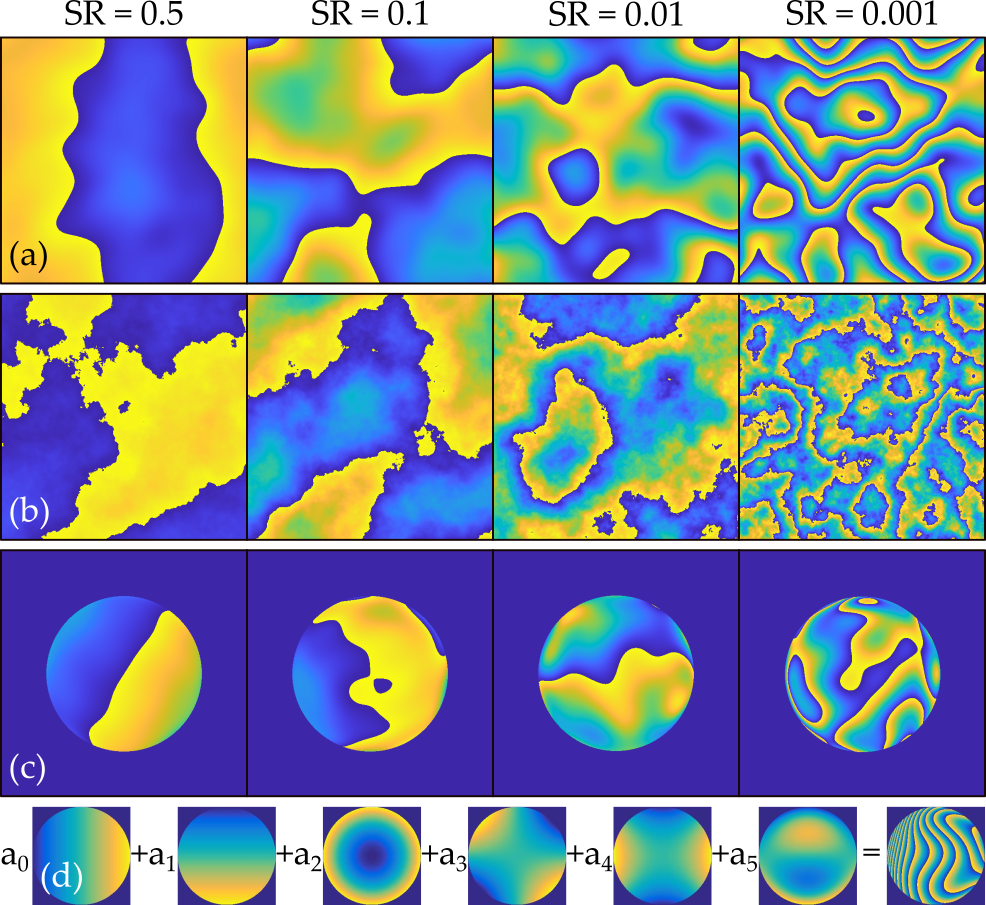}\vspace*{-0.5cm}
    \caption{Random phase screens of varying turbulence strength. Rows from top to bottom are for the naive inverse power spectrum approach (a), the sub-harmonic method (b) and the Noll (Zernike) method (c). The last row illustrates the individual Zernike phase screens which make up the Noll method (d), although in reality up to 60 screens are required for a physically accurate result. }
    \label{fig:TurbScreens}
\end{figure}

Examples of generated phase screens are shown in Fig.~\ref{fig:TurbScreens}~(a), in order of increasing turbulence from left to right. The screens as generated do not accurately model atmospheric turbulence since they cannot perfectly represent low-order spectral components and therefore produce statistics that do not conform to the theoretical structure function. As a result, sub-harmonic methods to include low order components have been developed \cite{Lane1992}. Example phase screens, that include these lower order components are shown in Fig.~\ref{fig:TurbScreens} (b). 

The second method generates phase screens, shown in Fig.~\ref{fig:TurbScreens}~(c), as superpositions of Zernike polynomials following the expansion
\begin{equation}
    \psi(r, \theta) = \sum_{j=0}^{J} a_j Z_{j}(r,\theta)
    \label{zernikeph}
\end{equation}
where $(r,\theta)$ are polar coordinates, $J$ is the last index of the used modes and $Z_{j}(r,\theta) \equiv Z_{m,n}(r,\theta)$ denote the Zernike polynomials. The Zernike modes are orthogonal over a unit circle and are commonly used to represent  wave-front distortions in optical beams. The modes can be written as 
\begin{align}
Z_{m,n} (r,\theta) =\left\{ \begin{array}{ll} 
                \sqrt{2(n+1)} R_{n}^{m}(r) G_m(\theta) & m\neq0 \\
                R_{n}^{0}(r) & m=0 
                \end{array} \right.,
\end{align}
\noindent where the index $j$ can be mapped onto the  double index, $(m,n)$, for positive integers $m\leq n$ following the Noll convention \cite{Noll1976}. The functions $R_{n}^{m}(r)$ and $G^m(\theta)$ are the radial and azimuthal factors given by 
\begin{equation}
R_{n}^{m}(r) =  \sum_{s=0}^{(n-m)/2} \frac{-1^s (n-s)!}{s! \ \left( \frac{n+m}{2} -s \right)! \ \left( \frac{n-m}{2} -s \right)!},
\end{equation}
\noindent and 
\begin{align}
G_m(\theta) = \left\{ \begin{array}{ll} 
                \sin(m\theta) & \hspace{5mm} j \ is \ \text{odd} \\
               \cos(m \theta) & \hspace{5mm} j \ is \ \text{even} 
                \end{array} \right..
\end{align}
Examples of the first few Zernike modes in the Noll convention are shown in Fig.~\ref{fig:TurbScreens}~(d), contributing in the construction of a phase screen. We list the first three indexes: $j=2$ ( or $(m,n)= (1,1)$), $j = 3$ (or $(m,n)=(1,1)$) and j = 4 (or $(m,n)=(2,2)$). Although the indexes $j=1$ and $j=2$ share the same index pairs, $(m, n)$, and therefore the same radial factor, they have differing phase factors due to the phase function, $G(\theta)$; one is an even function while the other is odd.

Further, for the Kolmogorov spectrum the coefficients, $a_j$, can be sampled from a normal distribution with zero mean and a variance of, $ \sigma_{nm}^{2}= I_{mn} \times \left( D / r_{0} \right) ^{5/3}$, where $I_{mn}$ can be determined from the diagonal terms of the Noll matrix \cite{Noll1976}
\begin{equation}
    I_{mn}=\frac{0.15337(-1)^{n-m}(n+1)\Gamma(14/3)\Gamma(n-5/6)}{\Gamma (17/6)^{2}\Gamma(n+23/6)}. 
\end{equation}
This produces phase screens with statistical properties consistent with the Kolmogorov model for atmospheric turbulence.

\begin{figure}[tb]
    \centering
    \includegraphics[width=1\linewidth]{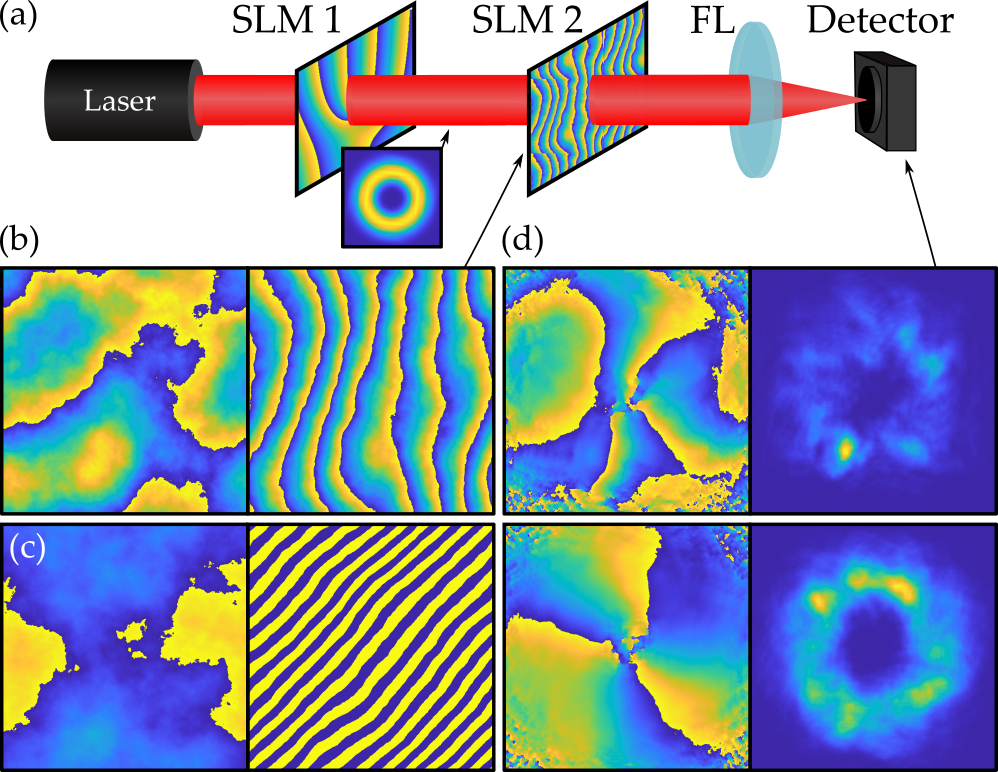}\vspace*{-0.5cm}
	\caption{\label{fig:turbScreensGeneration} (a) Typical set-up for creating a mode and then aberrating it with turbulence (FL: Fourier Lens; SLM: Spatial Light Modulator). (b) Holograms for strong (top, SLM) and weak (bottom, DMD) turbulence with and without gratings. (c) Phase and intensity of $\ell=3$ beams through the setup with the turbulence to the left.} \label{fig:TurbExamples}
\end{figure}


Once the phase functions corresponding to the phase screens have been generated, the next step is to produce holograms that can be used to imprint the desired perturbation onto an optical beam. By way of example we will demonstrate this procedure for SLMs and DMD devices.  Given the phase function, $\psi(x, y)$, a hologram tailored for a phase-only device, can be calculated from \cite{Rosales2017}  
\begin{equation}
 H (x, y) = \text{mod}\{ \psi(x, y) + k_{ox} x + k_{oy} y, 2\pi \}
\end{equation}
where the factors $k_{ox,oy}$ are the grating frequencies in the $x$ and $y$ (Cartesian) directions, respectively. In this way, a structured blazed grating, with a phase depth of $[0, 2\pi]$ is generated. Examples of this are shown in Fig.~\ref{fig:TurbExamples}, along with the resulting beam phase and intensity.

Some devices allow for a modulation of up to $8\pi$, which enables the simulation of steeper phase gradients with fewer pixels. The disadvantage of SLMs is their modulation speed which is typically tens of Hertz. The time scale at which turbulence changes is given by the Greenwood frequency, which for modest conditions is an order of magnitude higher. 

For this reason DMDs are ideal: they have modulation speeds of kilohertz, allowing accurate simulation of realistically fast changing turbulence. Unfortunately, DMDs use an amplitude only hologram encoding which limits efficiency: the first diffraction order of a DMD has $<10\%$ of the initial input power. The desired hologram can be calculated as \cite{scholes2019structured}
\begin{align}
 H (x, y) =& \frac{1}{2} + \text{sign} \Big( \cos \big[ k_{ox} x + k_{oy} y  + \pi p(x, y) \big] \\  &+  \cos( \ \pi w(x, y) \ ) \Big),
\end{align}
where $w(x,y) = (1 / \pi ) \times \arcsin( A(x, y) )$ and $p(x,y) = (1 / \pi ) \times \psi(x, y)$ corresponding to the amplitude and phase terms, respectively. Again, $\psi(x, y)$ plays the role of the phase function mapping the desired turbulence phase fluctuations. Since we only consider the phase variations, the amplitude term can simply be set to unity, i.e $A(x, y) = 1$. 

The resulting screens are binary images having zeros and ones as entries. We show an example of a DMD turbulence hologram for Strehl Ratio 0.9 (weak) turbulence in Fig.~\ref{fig:TurbExamples}~(c), along with the resulting beam phase and intensity.

Now that we generated some phase screens and are able to produce holograms that can be used to modulate optical beams, how do we know if they have imparted the desired turbulence strength? The simplest way is to  measure the Strehl Ratio (SR). By using Eq.~(\ref{eq:bsr}), it can be measured by comparing the on-axis intensity of the obstructed and unobstructed beams. The SR parameter produces a function scaling from 0 to 1, covering strong to weak turbulence, respectively.  It is instructive to exploit the relation between the normalised aperture size, $D/r_0$, and the SR in calibrating phase screens in the lab. Since we want to generate turbulence strengths with a specific coherence length, $r_0$, it is ideal to  calibrate the system by changing the aperture size argument, $D$, until the theoretical and experimental SR are equivalent for various values of $D/r_0$.

A schematic illustrating a typical setup for using this approach is shown in Fig.~\ref{fig:TurbExamples} (a) with examples of simulated and measured intensity patterns shown in Fig.~\ref{fig:TurbExamples} (b) and (c), respectively. In the setup, a collimated laser beam is incident on an SLM and imaged by a camera at the far-field of a Fourier lens for various turbulence strengths parameterised by $D/r_0$. 

\begin{figure}[tb]
  \centering
    \includegraphics[width=\linewidth]{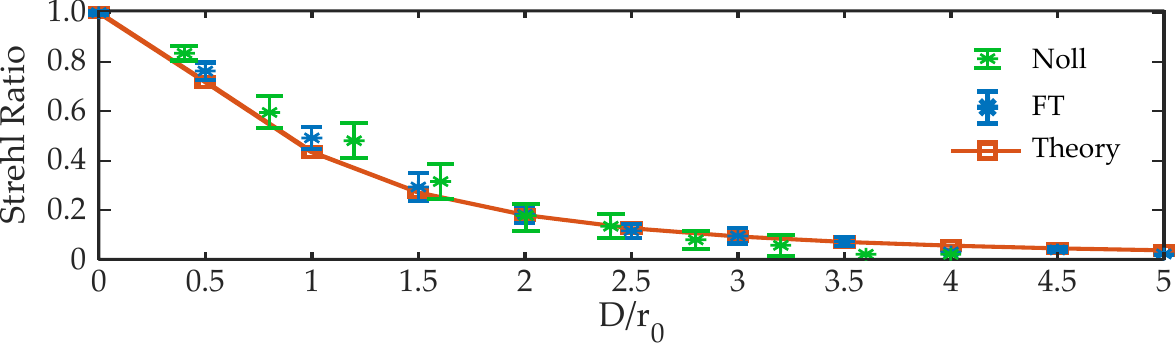}\vspace*{-0.6cm}
  \caption{Experimental measurement of Strehl Ratio using turbulence screens generated using the spectrum inversion (FT) and Zernike (Noll) methods showing a close match to theory.}
  \label{fig:srcalib}
\end{figure}

For each image, the on-axis intensity is recorded. The SR ratio can be calculated from these intensities and then compared to the theory.  Examples of the measured SR are shown in Fig.~\ref{fig:srcalib} for a well calibrated system using the sub-harmonic spectrum inversion and Zernike (Noll) methods. 


\subsection{Real-world turbulence}
In this section we move out of the laboratory and into the real-world, reporting measurements over a 150~m link to demonstrate the effect of turbulence on structured light in real time, exposing limitations of lab experiments using simulated turbulence screens.

\begin{figure*}[tb]
  \centering
      \includegraphics[width=\linewidth]{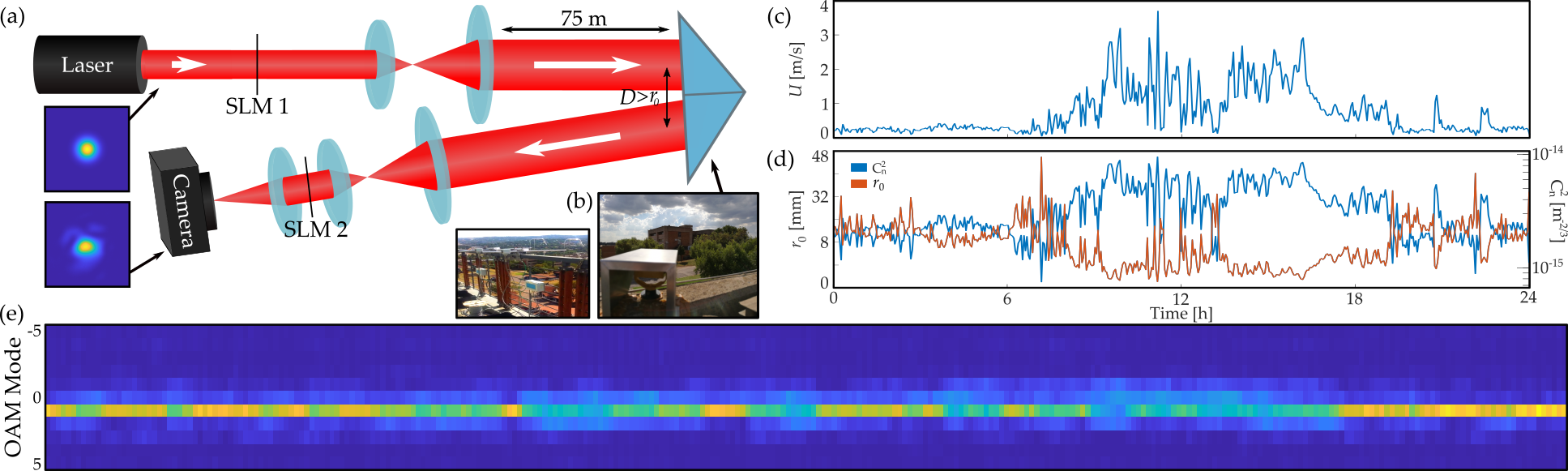}\vspace*{-0.5cm}
  \caption{(a) Experimental setup for the 150~m retro-reflector FSO link. (b) Photo of two sonic anemometers. (c) Measurements of mean wind velocity, $U$, used to calculate (d) $C_n^2$ and $r_0$ for $L= 150$~m over 24~hours using a sonic anemometer. (e) measurements of OAM crosstalk ($\ell=1$ transmitted) over one second at 300~Hz with visible ``flow'' evolution (See Sec.~\ref{sec:memory}).}
  \label{fig:exptsetup}
\end{figure*}

In a real-world experimental setup one of the challenges is accurately measuring the turbulence parameters to characterise the results. This is typically done by transmitting a Gaussian beam through a system such as that shown in Fig.~\ref{fig:exptsetup}~(a). The scintillation index, $\sigma_I^2$ given by Eq.~\ref{eq:bscintillationIndex}, can then be measured at the receiver using a camera or photodiode. If $\sigma_I^2 \ll 1$ then the link is in the weak turbulence regime and therefore $\sigma_I^2 = \sigma_R^2$. Using Eq.~\ref{eq:rytovCn2}, the $C_n^2$ value for the link can be found and from this many other parameters. Unfortunately, this technique is only valid in weak turbulence. If the technique is not possible, or the turbulence regime is not weak, it is possible to accurately measure $C_n^2$ using other weather parameters such as wind speed, temperature and pressure.  

To acquire this real-world data of the atmosphere we deploy a three-dimensional wind vector instrument, a sonic anemometer. A photo of two of these devices attached to the side of a building is shown in Fig~\ref{fig:exptsetup} (b). Here we will explain how to extract all the salient turbulence parameters to  understand the real-world link.  

A sonic anemometer is a solid-state, ultrasonic instrument which measures wind velocity in three orthogonal axes as well as the so-called sonic temperature. Sonic anemometers are commonly used for accurate wind measurements, but are sometimes used to measure near field turbulence around high towers, where optical communication links are situated. The instrument consists of three pairs of transducers separated by a 60 degree angle across the measurement volume. Each transducer in a pair acts alternately as either a transmitter or a receiver. The transducers are separated by a distance, $l$ (typically on the order of 10~cm) and the travelling time of the pulse in each direction is measured. 
%

These measurements are performed for each of the three transducer pairs simultaneously. A numerical transformation is then applied to convert the resulting three wind vector components, together with temperature and pressure measurements, into an instantaneous sonic temperature, $T_{s} (t)$, from which the time dependent temperature structure function constant, $C_T^2 (t)$, can be inferred.  Using the Gladstone-Dale law it is trivial to covert this to the desired refractive index structure function, $C_n^2$.  The advantage is that this device returns (by calculation) a time series of $C_n^2$ values, which we measured at intervals of three minutes over a 24~hour period, with the results shown in Fig.~\ref{fig:exptsetup}~ (c) and (d).

Figure~\ref{fig:exptsetup}~(c) shows measurements of the average wind speed over a 24 hour cycle, while Fig.~\ref{fig:exptsetup} (d) shows the derived refractive index structure function, $C_n^2$, of a daily course from observed values of the airflow, together with the $r_0$ value.  The derived $C_n^2 \approx 10^{-14} - 10^{-13} \text{m}^{-2/3}$ indicates moderate turbulence. It can be seen that scintillation values are lowest in the morning, highest around noon and decrease again after sunset. This is as expected, since there is less turbulence at night than during the day, primarily because there is much less atmospheric heating.  

To illustrate the real effect of turbulence on spatial modes, rather than just a simulation, an $L=150$~m horizontal FSO link was constructed at the same site (CSIR, Pretoria, South Africa), which is inland at an altitude of 1.4~km. A simplified diagram on the link is shown in Fig.~\ref{fig:exptsetup}. The link was approximately 5~m above the ground and used a $\lambda = 635$ nm collimated laser beam with a diameter of approximately $\omega_0 = 7$~mm at the transmit aperture. The transmitted and received laser beams were expanded and reduced by telescopes and imaged onto a high speed camera at the Fourier plane of a Spatial Light Modulator (SLM) which was used for modal decomposition experiments.  A wide, rectangular retro-reflector prism was used. There is an offset of several centimetres between the incoming and outgoing beams, as well as some beam deviation due to the precision of the prism, which was larger than the typical atmospheric coherence length of the turbulence at the time of the experiments. As such, atmospheric reciprocity (which negates beam wander) is not an issue in this experimental setup. 

\medskip

Figure~\ref{fig:exptsetup}~(e) shows a sample of OAM crosstalk measurements with a $\ell=1$ transmitted mode taken at 300 frames per second (or 3.33~ms between measurements). It is interesting to note that over time, the crosstalk appears to ``flow'' with a certain amount of memory. This phenomenon is not modelled in the literature, but we propose a methodology based on correlated angle-of-arrival fluctuations which can be used to accurately describe this in Sec.~\ref{sec:memory}.

Using the same link, we demonstrated the feasibility of shape-invariant Bessel modes for free-space optical communications \cite{mphuthi2019free}. These modes, previously generated in the laboratory, have longitudinally dependent cone angles which allow the Bessel structure to be preserved during propagation. This property overcomes the limit of traditional Bessel modes, which turn into an annular ring outside the specified depth of focus. Crosstalk measurements were also obtained through modal decomposition at different times of the day which constitute different turbulence conditions. The results were compared with helical Gaussian modes of the same OAM values, discussed further in Sec.~\ref{sec:robustness}.

\vspace{0.5cm} 
\section{Perspectives}
\label{sec:results}


What is the present state-of-the-art with respect to structured light in turbulence? Are there any families of structured light that are more robust than others? How much information can we pack into the spatial modes of light for a high-bandwidth FSO communication link? Is there gain to be had by introducing a turbulence invariant degree of freedom, polarisation, into the structure of light? Can we benefit from turbulence in any way? And how many of these classical approaches can be translated to the quantum realm to realise fast and secure FSO links?  In this section we review the present status of structured light in turbulence with particular emphasis on these universal questions. 

\subsection{Multiplexing versus diversity}

The drive towards MDM is fueled by the impending ``capacity crunch'' \cite{Winzer2014a},  where we are rapidly approaching the Shannon limit for information capacity in fibre. The space degree of freedom, unlocked in a compact manner by the use of orthogonal OAM modes is indeed a possible solution.  The first use of OAM in MDM was in 2004 \cite{Gibson2004}, and since then steadily advanced \cite{Willner2017,trichilliSDMreview}, gaining in popularity after a successful 2.56~Tbps demonstration \cite{Wang2012}.  There are questions as to whether there is an advantage in using OAM modes over others, with suggestions that if only the LG azimuthal degree of freedom is used (i.e., OAM), then there is no capacity increase over simply using other MIMO techniques or the full basis \cite{Zhao2015,Chen2016a}. Indeed, it has been argued that in order to increase the capacity of a link, we should simply perform a singular value decomposition of the channel and use whichever modes are best \cite{Miller2017}. While being optimal in a theoretical sense, this is not necessarily a practical solution because of the technical difficulty and cost of doing so. While SLMs or DMDs can be used to create arbitrary (and perhaps optimal) spatial modes, they are expensive and inefficient and much light is wasted. Instead, if a predetermined set of modes is used, compact, passive devices can be used such as meta-materials \cite{sephton2020purity} and diffractive optics \cite{Ruffato2018}.

Nevertheless, there has been significant progress and several impressive demonstrations in using structured light for multiplexing \cite{Huang2014,Milione2015,Zhu2016,Krenn2016,zhao2016linkdemo,Ren2016,Qu2016}. To date, the fastest MDM demonstration has sustained a data rate of 1.036~Pbps over a lab-bench using 26 OAM modes together with wavelength and polarisation \cite{wang2014fso1pb}. The farthest MDM FSO communication demonstration in turbulence has been 80~Gbps over a mere 260~m \cite{zhao2016linkdemo} - turbulence is clearly a significant challenge. 

\begin{figure}[tb]
    \centering
    \includegraphics[width=1\linewidth]{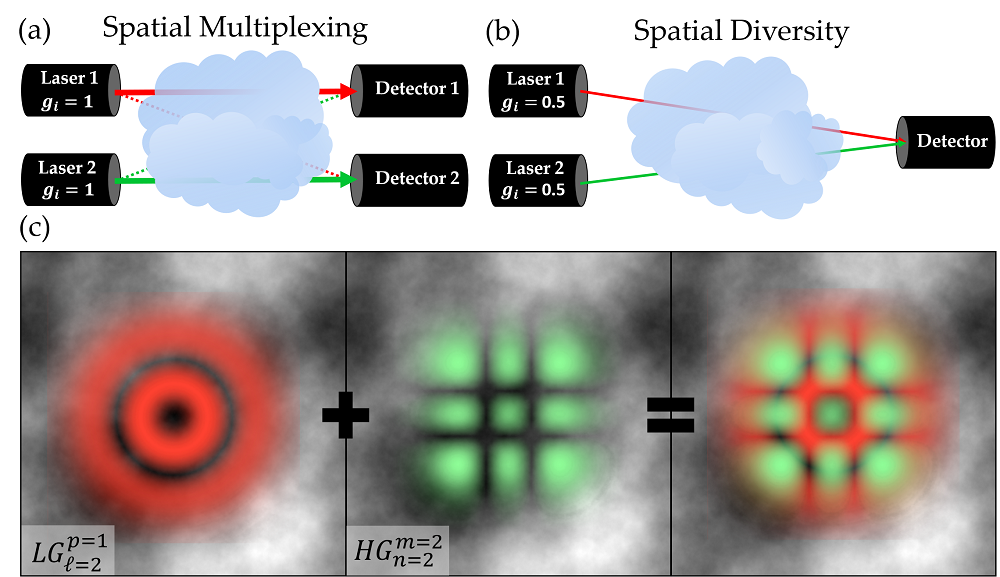}\vspace*{-0.5cm}
	\caption{\label{fig:mdmdiversity} The difference between multiplexing and diversity shown diagrammatically in (a) and (b). When structured light is used, the ``physical'' separation for multiplexing or diversity is in the form of orthogonal spatial modes rather than separated paths (c).  }
\end{figure}

In a system making use of the space degree of freedom, spatial multiplexing such as MDM is only one aspect of structured lights potential. It is common in FSO systems to use multiple transmit and / or receive apertures to increase the robustness of the system to turbulence induced fading, provided the apertures are separated by at least $r_0$ to ensure statistical independence, known as diversity. This is shown diagrammatically in Fig.~\ref{fig:mdmdiversity}. 

It has recently been shown that modal diversity also provides a so-called diversity gain, enabling a more compact system with fewer apertures \cite{Ark2014,Mehrpoor2015a,Huang2018}. Modal diversity here refers to the notion that differing spatial modes experience turbulence differently, and so have the capability of diversity: rather than use many modes for increased information content (multiplexing), one can use them to reduce error.  In particular, it was found that multiple modes with the same mode order can be used from different bases to also achieve a diversity gain, shown in Fig.~\ref{fig:mdmdiversity}~(c) \cite{cox2018diversity}. This was unexpected, as conventional turbulence theory does not predict independent aberrations of beams with identical size propagating together. This result supports the basis-dependent nature of turbulence discussed in Sec.~\ref{sec:robustness}. Subsequently there have been several experimental demonstrations using multiple mode bases and/or indices to increase the resilience to turbulence \cite{Zou2017,Pang2018HGLG,Amhoud2019}. 

The use of different modal bases in a system for multiplexing and diversity is an exciting new avenue of research. Using orthogonal modes with the same mode order (i.e., the same divergence properties) is particularly useful as it limits the required aperture sizes, and therefore the size and cost of a system. Given that ``conventional'' degrees of freedom in a communication system such as wavelength and polarisation are also usable in FSO communication, the ultra-high capacity provided by MDM may not be necessary for FSO in the near future. On the other hand, the ability to use modes to increase the resilience (and therefore the range) of a FSO link is immediately beneficial.

\subsection{The robustness of structured light}
\label{sec:robustness}

In an MDM scheme it is the spatial structure of the light that is exploited, and so any deviation in this structure will affect the channel capacity.  Take for example the simple Gaussian beam.  Gaussian beams have long been used in FSO links, but pertinently, the mode itself is not detected.  Instead, the detectors merely collect light in ``buckets'' without any spatial mode discrimination, while the information is encoded through modulation schemes such as on/off keying, phase modulation and so on.  In these traditional links turbulence introduces loss, affecting the intensity of the signal received.

In contrast, the same Gaussian mode could be one of many spatial mode information channels in MDM. Now, turbulence affects the very nature of each mode, including the Gaussian, so that modal cross-talk is induced in addition to deleterious effects on the signal. The turbulence aberrated modes are no longer orthogonal to one another, breaking a core ingredient of MDM.  Is there a structured light family that is less affected by turbulence, where the orthogonality is preserved?  



\begin{figure}[tb]
    \centering
\includegraphics[width=1\linewidth]{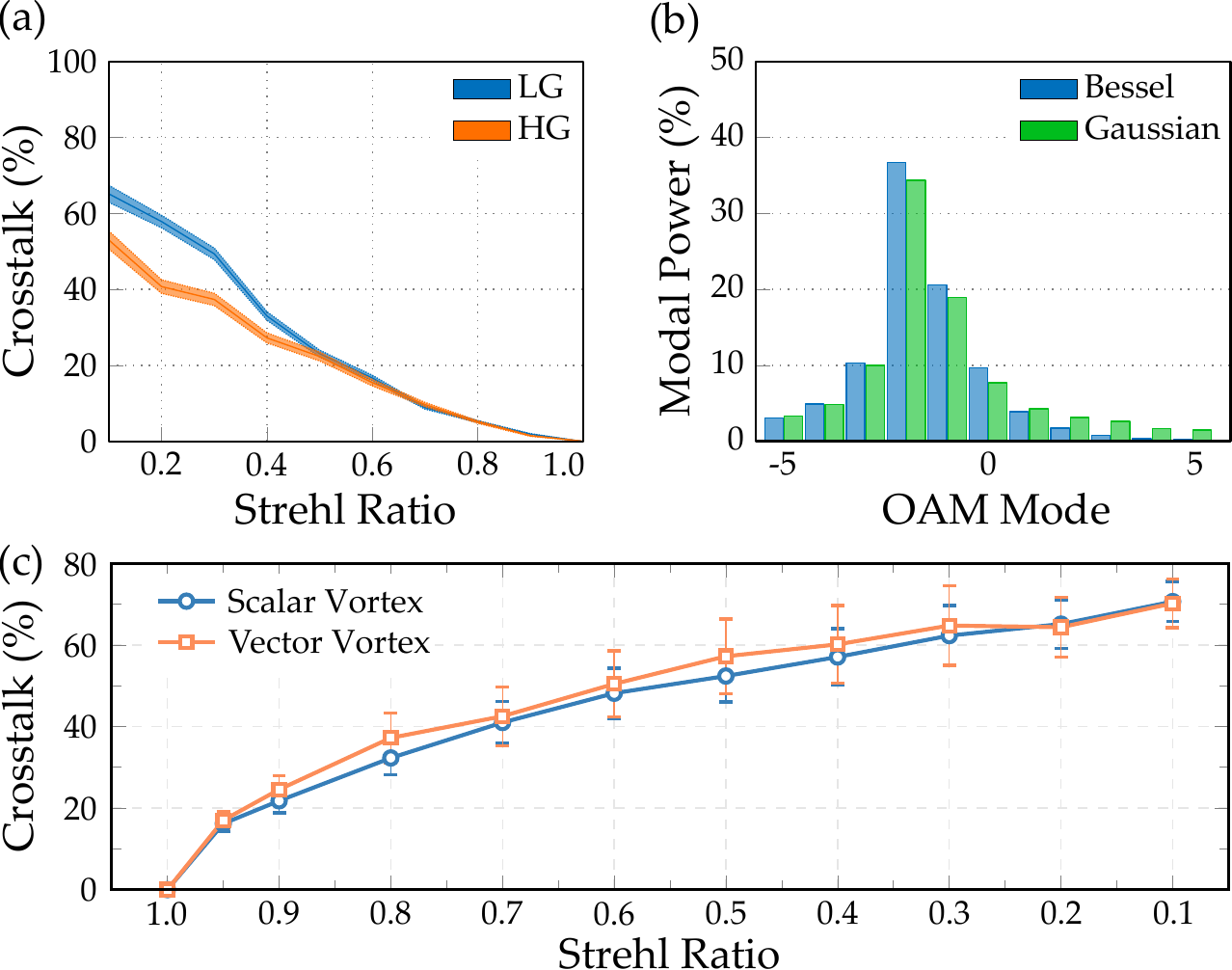}\vspace*{-0.5cm}
	\caption{\label{fig:lghgbg} (a) Crosstalk for HG and LG modes in turbulence characterised by Strehl Ratio showing that HG modes exhibit lower crosstalk in strong turbulence. (b) Long range Bessel modes versus a Gaussian in weak turbulence over 150~m showing lower losses for the Bessel beam. (c) Scalar versus vector vortex modes in turbulence showing identical performance. }
\end{figure}

Different sets of spatial modes have different characteristics and symmetries, which may result in basis-dependent performance in turbulence.  In support of this is the notion that Hermite-Gauss modes are more resilient to lateral displacements in comparison to OAM modes, owing to their Cartesian symmetry \cite{ndagano2017}. To understand this, recall that we used a weighted set of Zernike aberrations to describe turbulence.  From this expansion we note that tip and tilt, the angular beam directions in the horizontal and vertical, make up the majority of the effect of turbulence, which lead to beam wander in the far field.  The symmetry of low-order HG modes means that they are invariant to lateral displacements \cite{ndagano2017}. Even with the additional higher-order effects of turbulence, it has been experimentally shown that HG modes are more resilient than LG (or OAM) modes \cite{cox2019hglg}. An excerpt of the crosstalk results of LG versus HG modes in shown in Fig.~\ref{fig:lghgbg}~(a) where it is clear that at stronger turbulence strengths, HG modes (low order and excluding the symmetrical HG$^1_1$ case) are significantly more robust. 

On the other hand, it has been shown that in a perfectly aligned system with a restricted aperture, the channel capacity of a system using HG modes can be less than one using LG modes \cite{Restuccia2016}. This is because the circular shape of most optics suits the symmetry of an LG mode, but is not optimal for an HG mode. This work does not account for the zero on-axis intensity of many LG modes, which will have a significant alignment-dependent impact on the received signal.  

Interestingly, in simulations, Ince-Gauss modes are shown be superior to both HG and LG modes \cite{krenn2019turbulence}, however, experimental results are required to verify this.

There are some studies that suggest that Bessel-Gauss modes are more resilient to turbulence effects when compared to other modes. It was demonstrated through a series of numerical simulations that high order BG modes will outperform LG modes of the same order in terms of channel efficiency and bit error rates, particularly through high levels of turbulence \cite{doster2016laguerre}. This can be attributed to the BG modes' resistance to spreading as it propagates. It is due to the same reason that efficiently generated zero order BG modes with no side lobes present an advantage over Gaussian modes in terms of power delivery \cite{nelson2014propagation,birch2015long}. Interestingly, this is not unique to zero order BG modes as it has been found theoretically that low order OAM modes may in fact be more resilient than Gaussian modes \cite{Aksenov2016a}. It should be noted that the advantages of BG modes through turbulence do not extend to self healing. It was determined in a recent study that the self-healing effects of Bessel beams present no advantage when the phase is disturbed, as is the case when the modes propagate through atmospheric turbulence \cite{mphuthi2018bessel}. To overcome this, a modified Bessel mode with self healing properties could be considered \cite{mphuthi2019free}.

\medskip

The aforementioned discussion was restricted to scalar modes, yet the use of vector beams for optical communications has attracted significant interest of late \cite{lavery2014space,milione2015using,milione20154,wang2016advances,zhang2016120,zhang201795,huang2014100}, where the extra degree of freedom (polarisation) is useful in increasing channel capacity. Numerical studies have suggested that vector beams are more resilient to atmospheric turbulence effects than scalar beams \cite{cheng2009propagation,liu2011investigation,farias2015resilience,wei2015experimental}. It was found that the modes will experience less scintillation and its polarisation characteristics persevered over longer propagation distances. This claim was further substantiated by an experimental study where information was encoded on the orthogonal polarisation states of vector modes propagated through turbulence \cite{zhu2019compensation}. It was demonstrated that encoding information on the spatial polarisation profile rather than the complex field profile preserved the information during propagation. Another study showed that this notion remains true even through severe turbulence and the amount of scintillation is inversely proportional to the topological charge \cite{eyyubouglu2016scintillation}. 

\begin{figure}[t]
    \centering
	\includegraphics[width=1\linewidth]{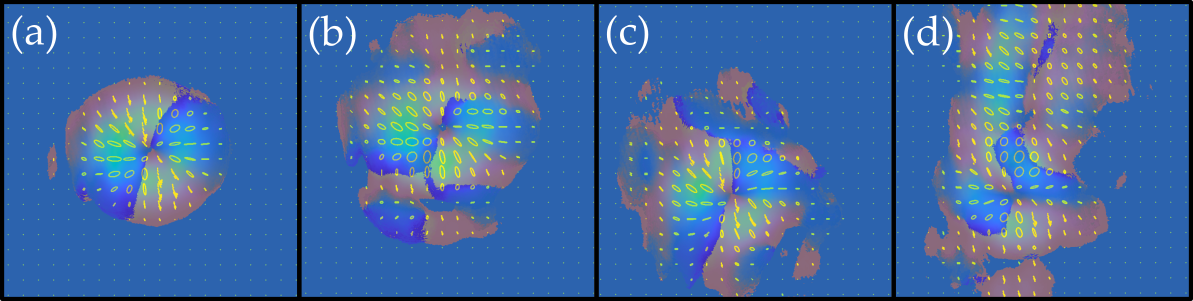}\vspace*{-0.5cm}
	\caption{Examples of vector modes propagated through thin turbulence phase screens. The polarisation, intensity  and phases maps are all superimposed. The vector modes are arranged in order of increasing turbulence strength from left to right.}
	\label{fig:vectmodeInTurb}
\end{figure}

By way of example, we illustrate the effect of turbulence on vector modes, showing superimposed polarisation fields, relative phases between the circular polarisation components as well as the intensity profile in Fig.~\ref{fig:vectmodeInTurb}. Each panel shows a vector mode under the effect of turbulence of varying strengths. By inspection, the intensity and polarisation profile of Fig.~\ref{fig:vectmodeInTurb}~(d) appears to be more distorted with respect to the undisturbed mode shown in Fig.~\ref{fig:vectmodeInTurb}~(a). However, if we ignore the lateral displacement of the beam (tip and tilt) in Fig.~\ref{fig:vectmodeInTurb}~(b) and (c), the modes maintain the characteristic polarisation singularity (null intensity and undefined phase) at the central (centre of mass) position of the beam. Intriguingly, the resilience of these modes seems to a have a strong bearing on the type of measurement technique used, for example, scintillation indexes and mutual information \cite{zhu2019compensation} in comparison to modal overlaps and fidelity measurements \cite{cox2016resilience} seem to yield conflicting results.

In addition, several studies have suggested that partially coherent vector modes have an advantage over completely coherent vector modes due to the lower scintillation index they experience under turbulence \cite{liu2013experimental,wang2013experimental,chen2013statistical,wang2014intensity,Chen2014}. This partial coherence coupled with polarisation modulation can also reduce spreading and beam wander effects allowing the modes to preserve the OAM information for much long distances \cite{cheng2019enhanced}. The advantages of vector modes also extends to the type of polarisation singularity present in the mode, where C-point singularities have been suggested to be more robust under turbulence than V-point singularities \cite{lochab2017robust}. 

However, there are cases where vector modes present no significant advantage over their scalar counterparts when propagated through atmospheric turbulence. It was shown \cite{cheng2009propagation} that the vortex structure of both scalar and vector OAM modes seem to disappear as the modes propagate through atmospheric turbulence. A theoretical study which investigated the irradiance and polarisation properties of cylindrical vector vortex (CVV) modes through turbulence showed that the vortex structure will evolve to a circular Gaussian spot, while the polarisation structure is completely destroyed \cite{cai2008average}. The same is true when inter-modal crosstalk is considered. Recent theoretical and experimental work showed that there is no performance benefit in using vector modes over scalar as the crosstalk due to turbulence is identical for both mode sets \cite{cox2016resilience}. This comparison of the crosstalk measurements of scalar and vector vortex modes under different turbulence strengths, quantified by Strehl ratio, is illustrated in Fig.~\ref{fig:lghgbg}~(c).


\subsection{Correction schemes}
All forms of structured light appear to be adversely affected by turbulence in some manner or other, and in these instances it is beneficial to compensate for the effects of atmospheric turbulence.
Several mitigation techniques have been employed in the compensation of scalar modes distorted by atmospheric turbulence and these include making use of adaptive optics \cite{zhao2012aberration,zhao2014turbulence,li2014evaluation,ren2014adaptive,ren2015turbulence,chen2016demonstration,zhao2016both,li2017adaptive,yin2018adaptive,li2019phase}, iterative approaches \cite{zhao2012aberration,li2017gerchberg,li2019phase} and most recently deep learning algorithms \cite{li2017adaptive,liu2019deep}. While some comparison studies of the different techniques suggest that adaptive optics will outperform iterative approaches when compensating for distorted OAM beams \cite{li2019phase}, some studies recommend a combination of different techniques to boost the performance of free-space communication links \cite{dedo2020oam,li2020atmospheric}. 
Significantly less work has gone into similar techniques and tools for vector beams.  Recent studies have looked into the correction of polarisation errors in vector modes but the application to atmospheric turbulence compensation is still in its infancy  \cite{chen2017demonstration,dai2019adaptive,dai2019active,de2019real,hu2020arbitrary,he2020vectorial,zhai2020turbulence}. 

It is topical to use computational power in the form of machine learning and Gerchberg-Saxton (GS) algorithms to find and correct for wavefront aberrations, first demonstrated with OAM beams over a decade ago \cite{jesacher2007wavefront}. The benefit of this approach stems from the fact that the phase information can be retrieved from only the measured intensity on a camera, without the need for wavefront sensors or reference waves required in interferometric techniques. Since there are many phase solutions that could result in the same intensity distribution as seen on the camera. Since the early seminal work, the approach has been used with scalar vortex beams as the probe to correct optical aberrations on DMDs \cite{scholes2019structured}, optical trapping systems \cite{baranek2018optimal} and OAM-based optical communication systems \cite{zhao2012aberration,ren2012correction}.

The technique is particularly sensitive when the test beam has a phase vortex \cite{jesacher2007wavefront}.  Here we point out that the vortex need not be phase in nature but could be a polarisation vortex, and so can be used to measure and correct aberrations on vector beams.  To demonstrate this, a radially polarised cylindrical vector mode was generated by superimposing two scalar vortex modes ($\ell=1$ and $\ell=-1$) using a SLM and an interferometer and sent through a DMD encoded with Zernike aberrations of unit variance. We demonstrate the concept here using an astigmatism aberration due to its detrimental effect on focal structure and rotational symmetry of the modes. 

\begin{figure}[tb]
\centering
\includegraphics[width=1\linewidth]{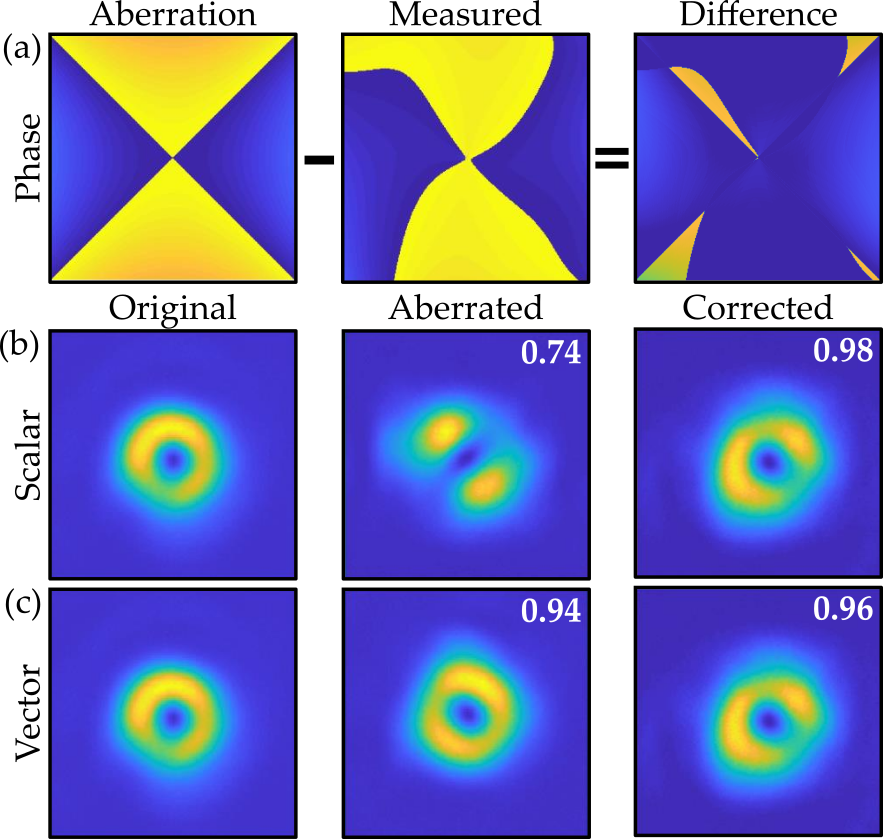}\vspace*{-0.5cm}
\caption{Experimental results illustrating the correction of astigmatism on scalar and vector beams. The aberration phase applied and measured (from the intensity using the GS algorithm) is shown in (a) with correction resulting in a nearly flat wavefront. The intensity profiles of the scalar (b) and vector (c) beams before (middle) and after corrections (right) are depicted with cross-correlations of the beam with the original beam (left).}
\label{fig:corrections}
\end{figure}

For each aberration, a camera image of the distorted vortex beam was captured and input on the GS algorithm to retrieve the wavefront errors. Phase vortices were used to retrieve errors from scalar modes while polarisation vortices were used to retrieve errors from vector modes. The retrieved phase was subtracted from the encoded aberration hologram to correct the mode. The mode fidelity was then calculated by performing a correlation measurement between an unaberrated mode and the corrected mode. The results were compared with scalar vortex beams aberrated by the same aberrations. Fig.~\ref{fig:corrections} displays the results.

A comparison between the encoded and retrieved phase show a close similarity with minor deviations. The phase was retrieved in about 30 iterations. The features associated with yhe aberration can be seen distinctly on the mode, represented by an elongation of the vortex structure. When observing the intensity profiles in Fig.~\ref{fig:corrections}~(a), we can see that there is an improvement in mode quality for both scalar and vector cases. For the scalar modes, the mode fidelity improved from 0.74 before corrections to 0.98 after corrections. 

The aberrated vector modes also seem to have a higher mode fidelity than their scalar counterparts even before any correction is applied. The mode fidelity for the aberrated vector beam before correction was 0.94 compared to 0.74 for scalar beams. This suggests that they are somewhat resilient to the aberrations. This trend seems to persist with each iteration step as the correction is applied (not shown). 

While an improvement in both scalar and vector beams is clear, we also notice that the increase for scalar modes is more significant than the vector case. This is due to the fact that the aberration shows a more distinct effect on the scalar vortex (such as lobe structure caused by astigmatism), which is recognisable by the algorithm and thereby able to retrieve the wavefront accurately.The overall fidelity of vector modes is, however, still more than that of scalar modes. This demonstrates the resilience of vector modes to aberrations which can also be attributed to the independence of each polarisation state which comprise the mode.

\subsection{Quantum structured light}

Quantum communication with  high  dimensional spatial  modes of light has received much interest in the quantum optics community with hopes of playing a substantial role in the future of global quantum networks to be deployed in our all-pervading noisy atmosphere \cite{forbes2019quantum, erhard2018twisted, roadmap, erhard2017quantum, PhysRevLett.123.070505}. Here the idea is to use structured light as single photon and entangled states to realise high-dimensional quantum states. While turbulence might inhibit the full potential of spatial mode encoding, there are several advantages: high dimensional states promise heightened security \cite{nagali2010cloning, bouchard2017high}, higher error thresholds \cite{mafu2013higher, Mirhosseini2015} and robustness against decoherence effects \cite{ecker2019overcoming}. 

To understand the impact of turbulence, a myriad of theoretical and experimental studies have been conducted on its impact on OAM entanglement \cite{paterson2005atmospheric ,roux2011infinitesimal, roux2015entanglement,leonhard2015universal, bachmann2019universal, jha2010effects, malik2012influence, ndagano2017characterizing}. The results indicate that OAM quantum states cannot withstand decoherent mechanisms due to turbulence. Despite this, they also highlight interesting features that can be of benefit. In particular, slower decay in quantum correlations has been predicted and confirmed for entangled OAM qubits with larger mode separation \cite{ibrahim2013orbital}.  In order for spatial modes to have any relevance, their high dimensionality has to come into play. Indeed, there has been substantial progress in this direction, with results confirming the robustness of high dimensional spatially encoded photons against turbulence in theoretical studies \cite{brunner2013robust} and on experimental test benches using dimensionality measures \cite{pors2011transport} and quantum state tomographies \cite{zhang2016experimentally}. In addition, methods for mitigating turbulence effects using high dimensional OAM states with adaptive optics \cite{sorelli2019entanglement, leonhard2018protecting, Zhao2020atmospheric} and effective channel correction \cite{mabena2020compressive}  exploiting compressive sensing techniques \cite{Ahn2019adaptive, mirhosseini2014comressive}, have also surfaced, therefore adding to repertoire of tools to be used for future realisations of high dimensional quantum communication in through turbulence.

Though we have discussed various attempts at studying spatial modes of photons within controlled environments, we now turn our attention to current state of the art in terms of real world applications. So far, there have been 4 successful attempts at realising long distance quantum communication with spatial modes in real world turbulence affected environment.

In the first demonstration \cite{vallone2014free}, the authors reported quantum encryption using rotation invariant qubits, over a 210 m outdoor link, using polarisation and spatial mode encoding. Here the authors did not use a true single photon source. The experiment was conducted under the effects of weak turbulence and showcased the benefit of rotation invariance for alignment free quantum communication. Following this achievement, the first demonstration of entangled spatial mode transport, over a 3~km link, was reported later \cite{krenn2015twisted}. The experiment was performed under strong turbulence conditions. Although the demonstration does not exploit the full potential of high dimensional encoding, it is however the longest spatial mode based quantum link reported to date. In another experiment, a group demonstrated high dimensional quantum cryptography with heralded single photons. The photons were encoded with four dimensional hybrid spin-OAM states, over a 300~m intra-city link \cite{Sit2017}. In their demonstration, the advantages of using high dimensional encoding were demonstrated in two ways; to increase the encoding capacity and to provide higher error thresholds. The experiment was conducted under moderate turbulence conditions with a structure constants in the range of $C_n^2 = 2.5 \times 10^{-15}$~m$^{-2/3}$ to $C_n^2 = 6.4 \times 10^{-16}$~m$^{-2/3}$. 
 
While progress has been made in realising real world quantum communication with spatial modes, to date there has been no successful attempt at transmitting high dimensional spatial mode entanglement through real world turbulence, which remains an open challenge. 

\subsection{The memory of turbulence}
\label{sec:memory}

Clearly the effect of turbulence on structured light is significant, and much work as been put into studying and mitigating it's effects. As mentioned in Sec.~\ref{sec:turb}, beam wander and angle-of-arrival fluctuations are the dominant cause of MDL and mode crosstalk. Existing models for these phenomena are memoryless and probabilistic in nature and do not describe their evolution with time.

It is known that the turbulence-induced fluctuations in the intensity of a received optical signal are correlated in time over several milliseconds due to the temporal behaviour of turbulence \cite{Davis1996,Rodriguez2005,Denic2008,Vasic2010,Anguita2010,Yura2011}. Various physical models have been proposed, with the simpler making use of Taylor's frozen flow hypothesis as well as complicated fractal models such fractional Brownian motion \cite{Zunino2014}. 

This memory behaviour is not accounted for in typical optical communication models for turbulence induced fading. Popular fading models, for example the recent double generalised gamma-gamma model \cite{Kashani2015}, are able to predict the expected amount of time a channel will be in deep fade over the long term, but inherently cannot provide information about the channel memory, which in this case is the duration of deep fade intervals or the crosstalk dynamics. This information informs critical design parameters in a communication system such as frame length, interleaver depth and error correction code length, and in it's absence researchers and engineers will continue to assume a memoryless channel, resulting in sub-optimal range or capacity. Furthermore, in a link using structured light, there are no models which allow us to predict how the crosstalk will evolve, except over a long term average. This hampers the development of crosstalk mitigation strategies. This is discussed further in Sec.~\ref{sec:dsp}.

Here we propose a novel approach to modelling focal spot wander due to angle-of-arrival fluctuations, which finds use in modelling and predicting the evolution of MDL and crosstalk. Our approach is based on the hypothesis that the movement of the wandering beam is not a true random walk, as is assumed in conventional angle-of-arrival and beam wander models, but rather a random walk that has a level of correlation between successive samples. We do not attempt to model the underlying physical phenomena, but rather focus on their effect, which is of significance to FSO communication systems.

\begin{figure}[tb]
    \centering
	\includegraphics[width=1\linewidth]{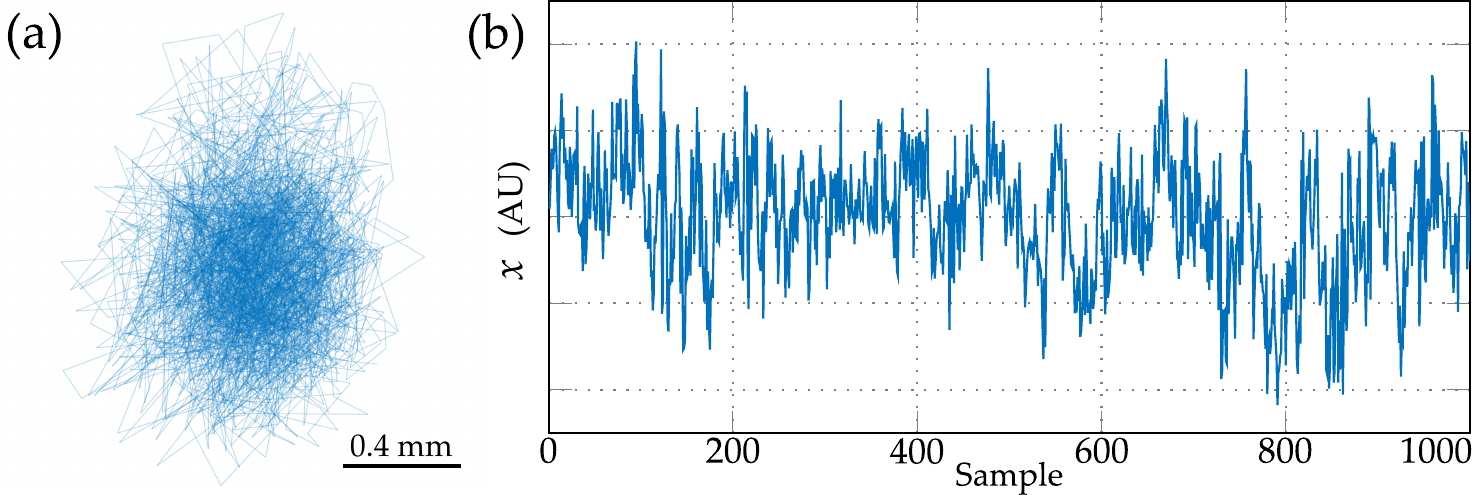}\vspace*{-0.5cm}
	\caption{\label{fig:wanderResults} Focal spot beam wander measurement (a) and the $x$-axis centroid position in arbitrary units over time (b). Each sample is 3.3~ms.}
\end{figure}

Measurements were performed using a Gaussian beam sent over the 150~m link described earlier. A high speed camera was used to measure a wandering beam at the focus of a lens, corresponding to the angle-of-arrival fluctuations at the receive aperture. The frame rate of the camera is an important parameter in this experiment as the movement of the beam must be recorded without loss of information.  From the environmental conditions (clear weather, 28~$\degree$C with an average wind speed of $v=10$~km/h with gusts of up to $v=12$~km/h) we find a Greenwood frequency of $f_G \approx 130$~Hz. The camera frame rate was therefore conservatively set at 300~Frames Per Second (FPS) to capture the movement of the beam without loss of temporal information. Further, the measured turbulence parameters of $C_n^2 = 4.1\times10^{-13}$, $r_0 = 0.01$~m and $\sigma_I = 0.55$, indicate that the link was in the moderate turbulence regime.  The position of the beam in each frame was found using the weighted centroid and stored as separate time series of $x$ and $y$ coordinates, with a sample of position measurements shown in Fig.~\ref{fig:wanderResults}~(a). Since the atmosphere is isotropic and the axes are orthogonal and independent, we restrict analysis to a single axis for convenience. When both axes are required for simulation of the resulting beam wander model, we simply run two models independently.

There are several classes of model that are suitable for correlated time series, depending on the characteristics of the signal. For stationary signals such as beam wander a so-called ARIMA (Auto-Regressive Integrated Moving Average) model is elegantly suitable, and can be fit using the Box-Jenkins method \cite{box2015time}. Auto-Correlation and Partial Auto-Correlation Functions (ACFs and PACFs) are used to approximately determine the order and suitability of the various model parameters, but ultimately the model fits performed using least-squares regression are verified by minimising the Akaike Information Criterion (AIC) and Bayesian Information Criterion (BIC).

A long term trend in the data is not expected, since a constant turbulence strength is assumed over the short duration of the measurements. Consequently, an ARMA($p$,$q$) model can be used without any Integral terms, where in general the position of the current sample is given by
\begin{equation}
    \label{eq:arma}
    x_{t} = y_t = c + \sum_{i=1}^p M_i x_{t-i} + \sum_{j=1}^q N_j \epsilon_{t-j} + \epsilon_t,
\end{equation}
where $c$ is a constant mean term, $p$ is the order of the auto-regressive series, $q$ is the order of the moving average series and $M_i$ and $N_j$ are the model factors of their respective series. Randomness is introduced to the model by the addition of $\epsilon_t$ which is zero mean Gaussian white noise with a suitable variance, $\sigma^2$. 

It was found that $x_t$ (and by nature $y_t$) is well modelled by an ARMA(2,2) process with the model factors shown in Tab.~\ref{tab:arma}, which were found using a least squares regression on the approximately 3000 measurement samples. The order of the model was verified by testing orders from zero to twenty and simultaneously minimising the AIC and BIC criterion's. Another test of the resulting model is given by the statistics of the residuals, which resemble random noise but are not shown here due to their noisy nature.

\begin{table}[tb]
\centering
\caption{Weights for the ARMA(2,2) model. As expected, the constant term is zero.}
\begin{tabular}{cccccc}
 $c$ & $M_1$ & $M_2$ & $N_1$ & $N_2$ & $\sigma^2$  \\
 \hline
 0 & 1.759 & -0.7626 & -1.289 & 0.3166 & 2150 \\
\end{tabular}
  \label{tab:arma}
\end{table}

As Eq.~\ref{eq:arma} is directly modelling the position of the beam at a specific time we are conveniently able to map this to a receive intensity. By assuming a short-term Gaussian beam with size $\omega$, the received intensity at a moment in time is given by
\begin{equation}
    \label{eq:mapping}
    I_t = I_0 \exp\!\!\left[-2\frac{x_{t}^2 + y_t^2}{\omega^2} \right],
\end{equation}
where $I_0$ is the intensity at the centre of the beam and the variables $x_t$ and $y_t$ are from two instances of Eq.~\ref{eq:arma}. Assuming that this model is accurate (as will be shown shortly), $I_t$ may be used as a fading factor in a Monte-Carlo simulation of an FSO link, and will provide accurate results which exhibit memory (i.e. accurate simulation of extended deep fading periods). This particular model does not include scintillation due to small scale turbulence, but this could be added as an additional stochastic factor.

\medskip

\begin{figure}[tb]
    \centering
	\includegraphics[width=1\linewidth]{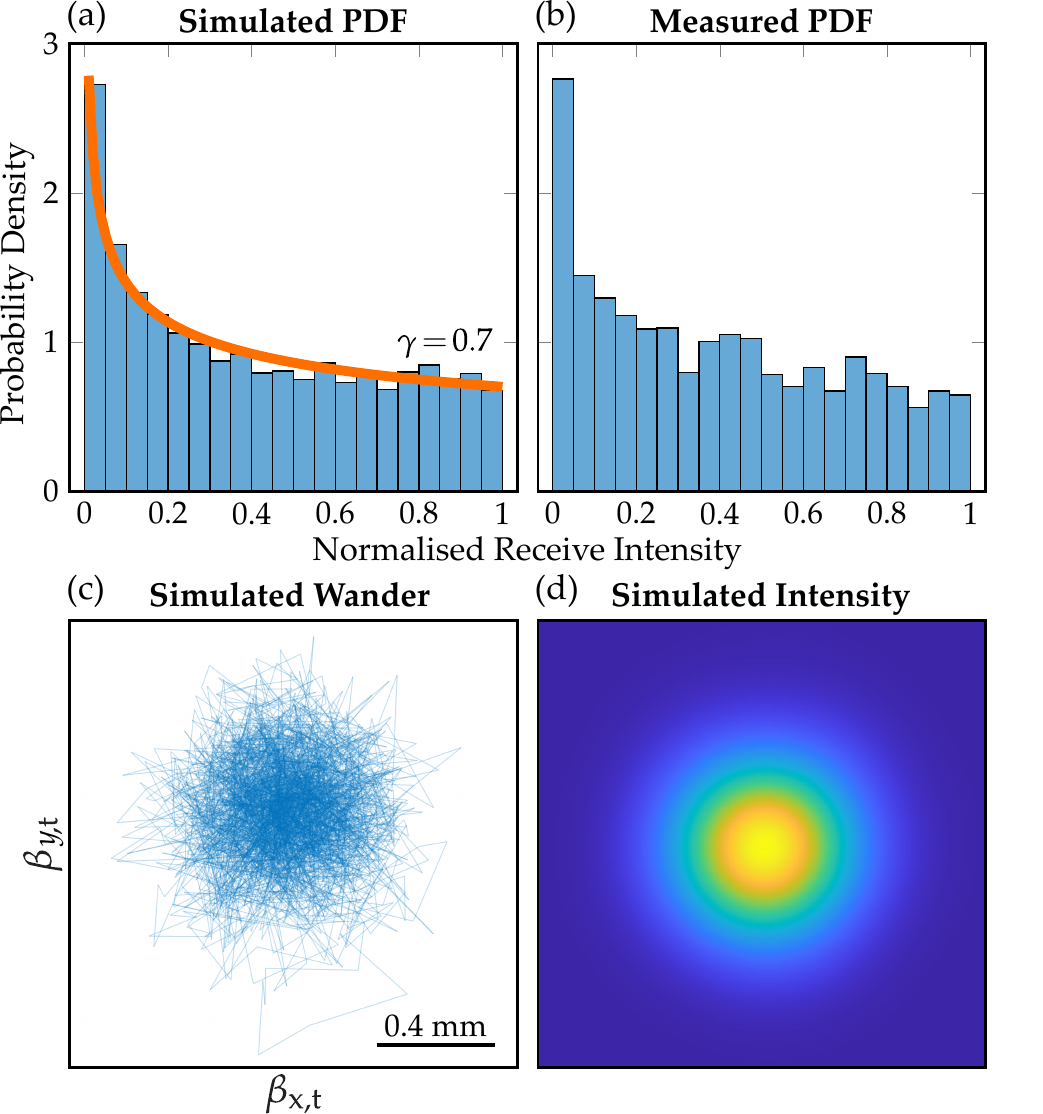}\vspace*{-0.5cm}
	\caption{\label{fig:simWander} Plots to demonstrate the fading statistics of the model in comparison to experimental data. (a) and (b) are PDFs of simulated and measured intensities with an overlay (orange) of the theoretical PDF showing excellent agreement with existing theory. (c) Simulated beam wander positions with the resulting long-term intensity in (d).}
\end{figure}

In order to verify this particular model, and hence this new approach, we must show that it is able to reproduce the results of existing models for beam wander. For a fair statistical comparison to the experimental data, the same number of samples of this ARMA model are generated as the number of experimental data samples. Plots of the output are shown in Fig.~\ref{fig:simWander}, where the Probability Distribution Functions (PDFs) of the experimental and model data are also shown. 

As expected, the long term beam intensity follows a Gaussian shape. For a valid model, the shape of the simulated PDF must be similar to that of the measured data. We see that this is indeed the case and in Fig.~\ref{fig:simWander}~(a) an overlay of the expected PDF model is also shown. This PDF is given by $p(I) = \gamma I^{\gamma-1}$ for $0 \leq I \leq 1$, where $\gamma$ is a parameter related to the ratio of beam wander displacement and the receiver size \cite{Kiasaleh1994}.  It is clear that this new approach results in models that accurately reduce to existing models for beam wander, however, their real benefit is the ability to model the channels memory.

A Run Length Distribution (RLD) is a histogram of the number of occurrences of a certain run length above or below a specified threshold. Here we use a RLD to visualise the amount of time the wandering beam spends inside or outside a radial threshold which corresponds to a certain intensity. Logically, with more data and regardless of memory, longer transition run lengths (which have correspondingly lower probabilities) will begin to appear on the RLD. For this reason, we can only compare data sets with the same number of samples. For illustration we set the transition threshold to 50\% of the maximum. 

\begin{figure}[tb]
    \centering
	\includegraphics[width=1\linewidth]{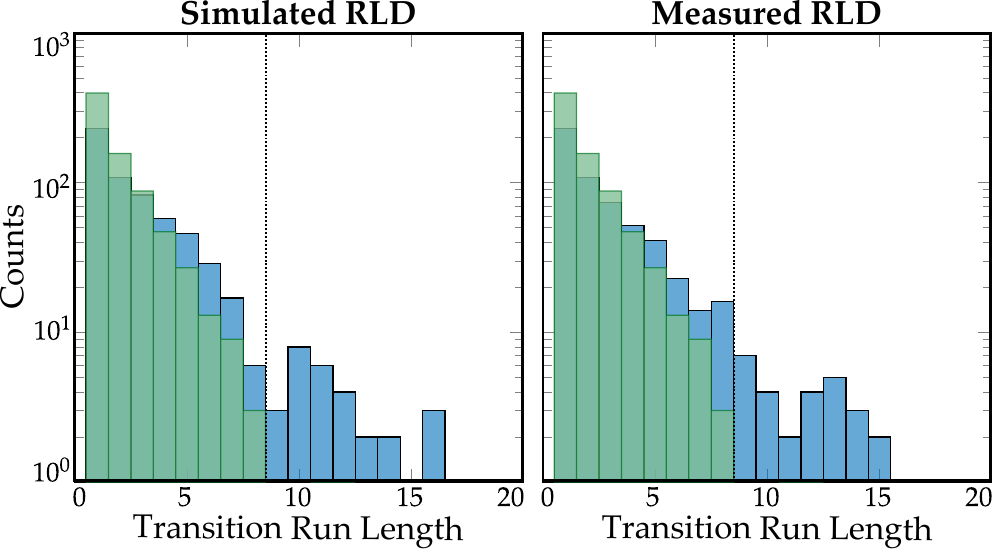}\vspace*{-0.5cm}
	\caption{\label{fig:runLength} Run length distribution plots of the simulated ARMA model and the experimental measurements with a transition threshold set at 50\% intensity for illustrative purposes. Overlaid on the plots in green is the run length distribution of data simulated using the conventional PDF-based model which clearly does not match the experimental measurements.}
\end{figure}

Figure~\ref{fig:runLength} shows run length distribution plots for the ARMA simulated and experimentally measured beam wander induced fading data. The simulated data matches the experimental data well. Superimposed on the figure is a run length distribution plot for simulated data based on the conventional PDF-based model for beam wander ($\gamma=0.7$), which is shown in Fig.~\ref{fig:simWander}~(a). The run length distribution of the conventional PDF-based model drops off steeply and is clearly not a match for the experimental data if we are concerned with the memory of the angle-of-arrival fluctuations.

\medskip

The proposed model has further utility than modelling received intensity fluctuations and their memory. Reliable and high speed long distance MDM systems are extremely challenging because of crosstalk issues. The computational complexity of strong error correction codes and MIMO is prohibitive at link speeds requiring the use of MDM in the first place, and a better model may enable the development of more computationally efficient algorithms (i.e., short cuts in the maximum likelihood algorithms, perhaps).

Using the modelling approach in this paper it becomes possible to simulate and even predict the temporal evolution of mode crosstalk due to angle-of-arrival fluctuations and perhaps even lateral displacement at the receive aperture. 

Here we briefly demonstrate this concept using LG mode crosstalk. A transmitted LG mode with $\ell=0$, which is a standard Gaussian, will result in a spectrum of detected OAM modes at the receiver. The detected OAM spectrum due to a focal spot wander at a time $t$ is given by  \cite{Lin2010oamtilt}
\begin{equation}
\label{eq:modeCrosstalkWander}
    C_{\ell,t} = \exp\left(-\frac{(2r_{a,t}/\omega_0)^2}{4} \right) I_{|\ell|}\left(\frac{(2r_{a,t}/\omega_0)^2}{4} \right)
\end{equation}
where $r_{a,t} = \sqrt{x_t^2+y_t^2}$, $C_\ell$ is the weight for the $\ell$th mode and $I_n(x)$ is the $n$th-order modified Bessel function of the first kind. 

\begin{figure}[tb]
    \centering
	\includegraphics[width=1\linewidth]{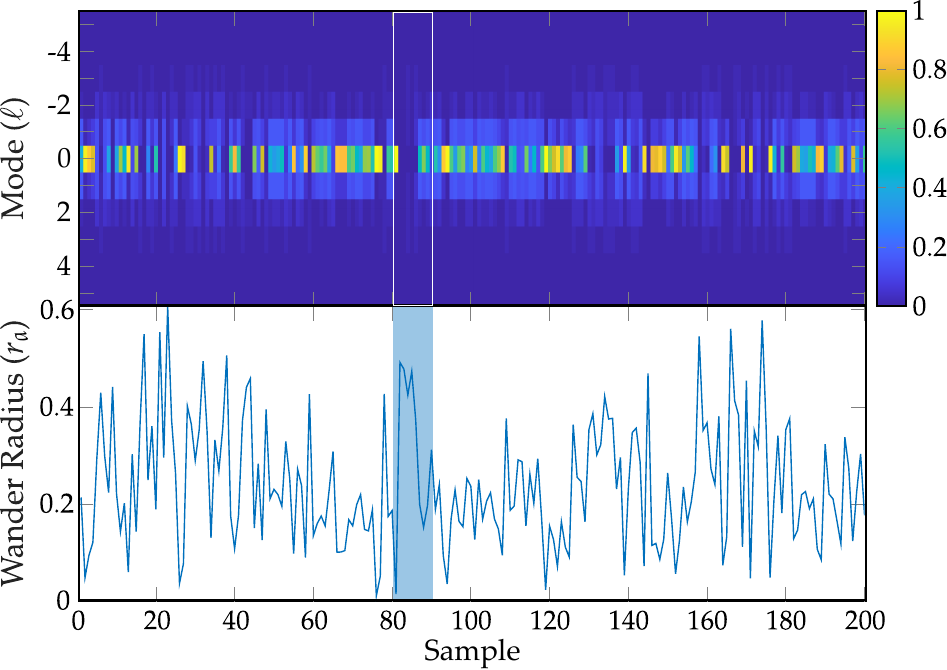}\vspace*{-0.5cm}
	\caption{\label{fig:wanderCrosstalk} (Top) Mode crosstalk from the transmitted Gaussian mode ($\ell=0$) into OAM modes $\ell=-5$ to $5$ due to angle-of-arrival fluctuations over time. (Bottom) The corresponding normalised focal spot beam wander radii, $r_a$. The highlighted section shows good examples of crosstalk with well aligned (minimal crosstalk) and misaligned (high crosstalk) cases.}
\end{figure}

Figure~\ref{fig:wanderCrosstalk} shows a sample of consecutive OAM crosstalk calculations from the simulated focal spot beam wander. Since we have shown a temporal correlation between beam wander positions, there will also be a similar temporal correlation between mode crosstalk spectrums. This may be harnessed in a predictive manner to implement more efficient signal processing strategies for MDM or modal diversity.

\section{Impact on optical signal processing}
\label{sec:dsp}


A key reason why communication errors due to turbulence is difficult to mitigate using signal processing and error correction coding is because of the significant probability that the received intensity will drop below a usable threshold \textit{for an extended period of time} - this is known as a ``bursty'' channel. The situation when the received intensity is too low for reliable communication is called a deep fade. In addition to this, the channel exhibits memory which causes the received intensity to remain in a certain state for a finite period of time. We provide a novel approach to modelling this in Sec.~\ref{sec:memory}. The combined effect of deep fading and memory is generally catastrophic to communication, where millions or even billions of bits can be lost.

Passive optical techniques such as transmit and receive diversity are very effective \cite{MIMOShin,MIMOKim}, but require the use of multiple apertures which increases the size and cost of a system. Modal diversity is a possible alternative which only requires one large transmit and / or receive aperture, but is in it's infancy \cite{Mehrpoor2015a,Huang2018,cox2018diversity}.

Generally, a diversity system is made more effective by the addition of some signal processing such as Equal Gain Combining (EGC) or Maximal Ratio Combining (MRC) \cite{MIMOGamma,Priyadarshani2018a}. Space-time block codes (for example Alamouti coding) have also been shown to be effective, but simple repetition codes (i.e. time diversity) may be more effective \cite{Simon2005a,Safari2008,Amhoud2019alamouti}. Repetition codes are a possible solution to dealing with channel memory, although they introduce latency which would be significant in a multi-hop FSO system, for instance. In addition, full MIMO processing has been shown to be effective especially when spatial modes are used, although is computationally complex especially at high data rates \cite{Ren2016b}. Forward Error Correction (FEC) also provides some error protection in FSO by adding redundancy to the transmitted bits \cite{ErrorCodingZhu,XuFEC09,Ren2012}. Since the FSO channel is typically slow fading and bursty, not all FEC codes are effective at combating the effects of turbulence. There has been some use of machine learning for mode detection, generally making use of OAM superpositions to create an identifiable ``petal'' structure, but these techniques have not yet been applied to communications (for instance in soft decision decoding) \cite{Park2018,Zhao2018,Krenn2016}.

Active optical techniques are typically very effective as they deal directly with the source of the problem. An issue, however, is that they are generally more expensive than digital techniques. For example, spatial light modulators, wavefront sensors and deformable mirrors are not as mass produced as DSP chips or field programmable gate arrays, and therefore not as cost effective. 

Since beam wander and angle-of-arrival fluctuations are dominant in turbulence, a simple tip/tilt mirror can be used as a rudimentary adaptive optics system, ensuring that the received beam is always aligned onto the detector or detection hologram \cite{Fernandez2018}. In addition, the receiver and / or transmitter can also be tilted to compensate for beam wander and pointing jitter \cite{Abadi2019}. Adaptive optics that also make use of deformable mirrors (used in astronomy) are very effective at correcting wavefront error introduced by turbulence, and work surprisingly well for OAM beams even though the systems are designed for Gaussian beams \cite{Ren2014}. It has also been shown that pre-distorting the transmitted beam so that it is corrected by the turbulence is effective \cite{Ren2014b,Almaiman2020}, however, this requires feedback in the system which consumes valuable bandwidth.

\section{Discussion and Conclusion}
Despite optical communication in turbulence being a topic with a very long history, it is receiving renewed attention in the context of structured light.  The spatial modes used for FSO communication are after all just orthogonal ``patterns'', but patterns become distorted in turbulence, adversely affecting the channel capacity.  Many studies have made headway in understanding the challenges, unravelling the performance of various modes in turbulence, considering correction schemes and even exploiting performance differences for diversity gain. But there remain many practical hurdles which has so far restricted the use of structured light in FSO links to kilometer scale distances.  

In this article we have reviewed the state-of-the-art in structured light in turbulence and introduced the core theoretical and experimental toolkit to advance the topic further.  We have used illustrative examples throughout the text, and provided a balanced account of the open questions in the community.  To add value to the tutorial style review we have used illustrative examples throughout, reporting in-house experimental results under a range of conditions.  In the process we have introduced some new results and insights, particularly those pertaining to vector modes in laboratory simulated turbulence, including an iterative approach to their correction, real-world turbulence results over a 150 m link for various beam types, highlighting the memory of turbulence and its impact on signal processing. 

The challenge is to take the many laboratory demonstrations into the real-world, a necessary step to push the limits in distance and information capacity.  We hope that this holistic article is a valuable resource to facilitate this.

\section*{Acknowledgments}
The authors would like to thank Angela Dudley and Lucas Gailele for the use of the the FSO link. M.A.C. acknowledges NRF grant 121908 and NLCRPP191024484931.

\ifCLASSOPTIONcaptionsoff
  \newpage
\fi


\bibliographystyle{IEEEtran}
\bibliography{bib.bib}
%



%

\begin{IEEEbiographynophoto}{Mitchell Cox}
received his PhD degree from The School of Electrical and Information Engineering at the University of the Witwatersrand, South Africa, in 2019 and also holds a MSc degree in Physics which he left industry to pursue. His research interests include improving the capacity and range of free-space optical communications using structured light, channel coding and signal processing. In 2019, Mitch received a University Vice-Chancellor award for his research citations. When Mitch isn't working with optics, he spends his time tinkering with electronics and software as an avid hobbyist while wondering how to make internet access a fundamental human right.
\end{IEEEbiographynophoto}

\begin{IEEEbiographynophoto}{Nokwazi Mphuthi}
is a PhD student with a MSc in Land Surveying from the University of KwaZulu-Natal in South Africa. As part of her MSc project, she worked on a project in collaboration with the South African Radio Astronomy Observatory (South Africa), NASA (USA), and Observatoire de la Côte d’Azur (France) to develop the first Lunar Laser Ranging (LLR) system in the Southern Hemisphere. An extension of the LLR work to increase the efficiency of the system paved the way for collaborations between the Structured Light lab at the University of the Witwatersrand and SARAO. The project aims to increase the photon return rate of laser ranging systems using structured light and someday provide means for sending orbital angular momentum to the Moon.
\end{IEEEbiographynophoto}


\begin{IEEEbiographynophoto}{Isaac Nape}
received  his  MSc  degree (with distinction)  from  the  School  of  Physics  at the University of the Witwatersrand, South Africa, and is currently a PhD student. His research interests include using spatial modes of light for quantum information encoding with single photons. To this end, his research outputs include several publications in the areas of quantum key distribution, quantum secret sharing and long distance quantum information transfer over optical fiber. 
\end{IEEEbiographynophoto}

\begin{IEEEbiographynophoto}{Nikiwe Mashaba}
received her Postgraduate Diploma in Nuclear science and technology from North West University, South Africa. She also holds a Diploma in Industrial physics and a Project management certificate from the University of Pretoria. She is currently completing her MSc degree in Physics at the Witwatersrand University and works for a research council, CSIR. Nikiwe's interests involves modelling and simulation of optical surveillance systems. 
\end{IEEEbiographynophoto}


\begin{IEEEbiographynophoto}{Ling Cheng}
(M’10-SM’15) received the degree BEng Electronics and Information (cum laude) from Huazhong University of Science and Technology (HUST) in 1995, M. Ing. Electrical and Electronics (cum laude) in 2005, and D. Ing. Electrical and Electronics in 2011 from University of Johannesburg (UJ). His research interests are in Telecommunications and Artificial Intelligence. In 2010, he joined University of the Witwatersrand where he was promoted to Full Professor in 2019. He serves as the associate editor of three journals. He has published more than 90 research papers in journals and conference proceedings. He has been a visiting professor at five universities and the principal advisor for over forty full research post-graduate students. He was awarded the Chancellor’s medals in 2005, 2019 and the National Research Foundation rating in 2014. The IEEE ISPLC 2015 best student paper award was made to his Ph.D. student in Austin. He is a senior member of IEEE and the vice-chair of IEEE South African Information Theory Chapter.
\end{IEEEbiographynophoto}

\begin{IEEEbiographynophoto}{Andrew Forbes}
received his PhD degree from the University of Natal, South Africa, in 1998. He subsequently spent several years as an Applied Laser Physicist, including in a private laser company at which he was the Technical Director and later as the Chief Researcher and the Research Group Leader of the Mathematical Optics Group, CSIR. He is currently a Distinguished Professor with the School of Physics, University of the Witwatersrand, South Africa, where he has established a new laboratory for structured light. He is a founding member of the Photonics Initiative of South Africa, a Fellow of SPIE and the OSA, an elected member of the Academy of Science of South Africa, and serves as Editor-in-Chief of the IoP's Journal of Optics.
\end{IEEEbiographynophoto}





\end{document}